\documentclass[aps, reprint, superscriptaddress, nofootinbib]{revtex4-1} 
\usepackage[dvipdfmx]{graphicx}
\usepackage{dcolumn}
\usepackage{amsmath}
\usepackage{amssymb}
\usepackage{color}
\usepackage{bm}
\usepackage{braket}
\usepackage{txfonts}

\begin{document}
\title{Oscillating-mode gap: an indicator of phase transitions in open quantum many-body systems}
\author{Taiki Haga}
\email[]{taiki.haga@omu.ac.jp}
\affiliation{Department of Physics and Electronics, Osaka Metropolitan University, Sakai-shi, Osaka 599-8531, Japan}
\date{\today}

\begin{abstract}
It presents a significant challenge to elucidate the relationship between the phases of open quantum many-body systems and the spectral structure of their governing Liouvillian, which determines how the density matrix evolves.
Previous studies have focused on the Liouvillian gap, defined as the decay rate of the most slowly-decaying mode, as a key indicator of dissipative phase transition, noting its closure in symmetry-broken phases and opening in disordered phases.
In this work, we propose an additional spectral gap, termed the oscillating-mode gap, defined as the decay rate of the most slowly-decaying oscillating mode.
Through the analysis of a prototype dissipative boson system, we demonstrate the necessity of both the Liouvillian gap and the oscillating-mode gap for the comprehensive characterization of the system's phases and the transitions between them.
\end{abstract}

\maketitle

\section{Introduction}

In quantum many-body physics, the spectral gap plays a pivotal role in defining the physical properties and phase behavior of the system.
Here, the spectral gap refers to the energy difference between the ground state and the first excited state of a quantum system.
It serves as a critical indicator of stability and quantum correlations within the system \cite{Hastings-04-1, Hastings-04-2, Hastings-06, Hastings-07}.
In particular, the closing of spectral gap signals quantum phase transitions (QPTs), marking points where the system undergoes dramatic changes in its ground state properties \cite{Sachdev}.
The behavior of spectral gap near the QPT point reflects universal properties characteristic of critical phenomena.

Recent research in open quantum many-body systems interacting with an environment has seen significant advancements, driven by breakthroughs made in realizing these systems experimentally, such as cold atoms \cite{Bloch-08-1, Bloch-08-2, Syassen-08, Bloch-12, Ritsch-13}, trapped ions \cite{Lanyon-09, Barreiro-11, Blatt-12}, and optical cavities \cite{Hartmann-06, Baumann-10, Eichler-14, Rodriguez-16}.
In general, the non-unitary evolution of the system's density matrix $\rho$ is described by the quantum master equation \cite{Lindblad-76, Gorini-76, Breuer, Rivas}:
\begin{equation}
\frac{d \rho}{dt} = \mathcal{L}(\rho) := -i[H, \rho] + \sum_{\nu} \left( L_{\nu} \rho L_{\nu}^{\dag} - \frac{1}{2} \{ L_{\nu}^{\dag}L_{\nu}, \rho \} \right),
\label{master_eq_general}
\end{equation}
where $H$ is the system's Hamiltonian and $L_{\nu}$ is a jump operator for dissipation.
The superoperator $\mathcal{L}$ is referred to as the Liouvillian.
A deeper understanding of coherent and incoherent processes modeled by the Liouvillian not only enhances our fundamental understanding of quantum many-body physics but also paves the way for practical applications in quantum computing and information processing \cite{Lanyon-09, Barreiro-11, Blatt-12}.

The concept of spectral gap extends into open quantum systems through the Liouvillian spectrum, which refers to the set of its eigenvalues.
These eigenvalues possess complex values, with the real parts corresponding to the decay rates of the respective eigenmodes, and the imaginary parts signifying their oscillation frequencies.
The Liouvillian gap, defined as the smallest absolute value of the real part among the non-zero eigenvalues, serves as a direct analogue to the energy gap in closed quantum systems.
It has been established that the closing of the Liouvillian gap signals a dissipative quantum phase transition (DQPT) within the steady state of the system \cite{Kessler-12, Honing-12, Horstmann-13, Casteels-16, Casteels-17, Fitzpatrick-17, Vicentini-18, Minganti-18, Imamoglu-18, Rota-18, Ferreira-19, Tomadin-11, Lee-11, Torre-13, Lee-13, Ludwig-13, Carr-13, Sieberer-13, Sieberer-14, Marcuzzi-14, Weimer-15, Maghrebi-16, Sieberer-16, Biondi-17, Domokos-17, Young-20}, thereby influencing its relaxation dynamics.

An open question in the study of open quantum many-body systems pertains to the adequacy of the Liouvillian gap for the comprehensive characterization of phases and transitions therein. 
Notably, the spectrum of the Liouvillian exhibits a two-dimensional structure within the complex plane, in stark contrast to the Hermitian systems, where the eigenvalues are real and thus the spectrum possesses a one-dimensional structure.
This fundamental distinction can necessitate a consideration of additional spectral gaps beyond the traditional Liouvillian gap to fully capture the dynamics and phase behaviors of open quantum many-body systems.
In fact, we will present a particular example of DQPTs occurring between phases characterized by a vanishing Liouvillian gap, illustrating the need of a paradigm that extends beyond the conventional framework for spectral characterization of phase transitions.

In the present study, we proposes an additional spectral gap for the Liouvillian, termed the oscillating-mode (OM) gap.
This gap is defined as the smallest absolute value of the real part among eigenvalues that possess a nonzero imaginary part.
Note that the OM gap is a feature unique to non-Hermitian systems which accommodate complex eigenvalues.
Leveraging the interplay between the Liouvillian gap and the OM gap, we propose a classification of the spectra for open quantum many-body systems into four distinct types, as illustrated in Fig.~\ref{fig_spectrum_class}.
Employing a mean-field approximation and exact diagonalization, our examination of prototype dissipative boson systems reveals the manifestation of such spectral types and elucidates the transitions occurring among them.
Our findings suggest a profound correlation between the spectral type of a system and its inherent symmetries.
Furthermore, we also discuss the impact of the OM gap's closure on the system's relaxation dynamics toward its steady state.

\section{Oscillating-mode gap}
\label{sec_oscillating_mode_gap}

\begin{figure}
\centering
\includegraphics[width=0.45\textwidth]{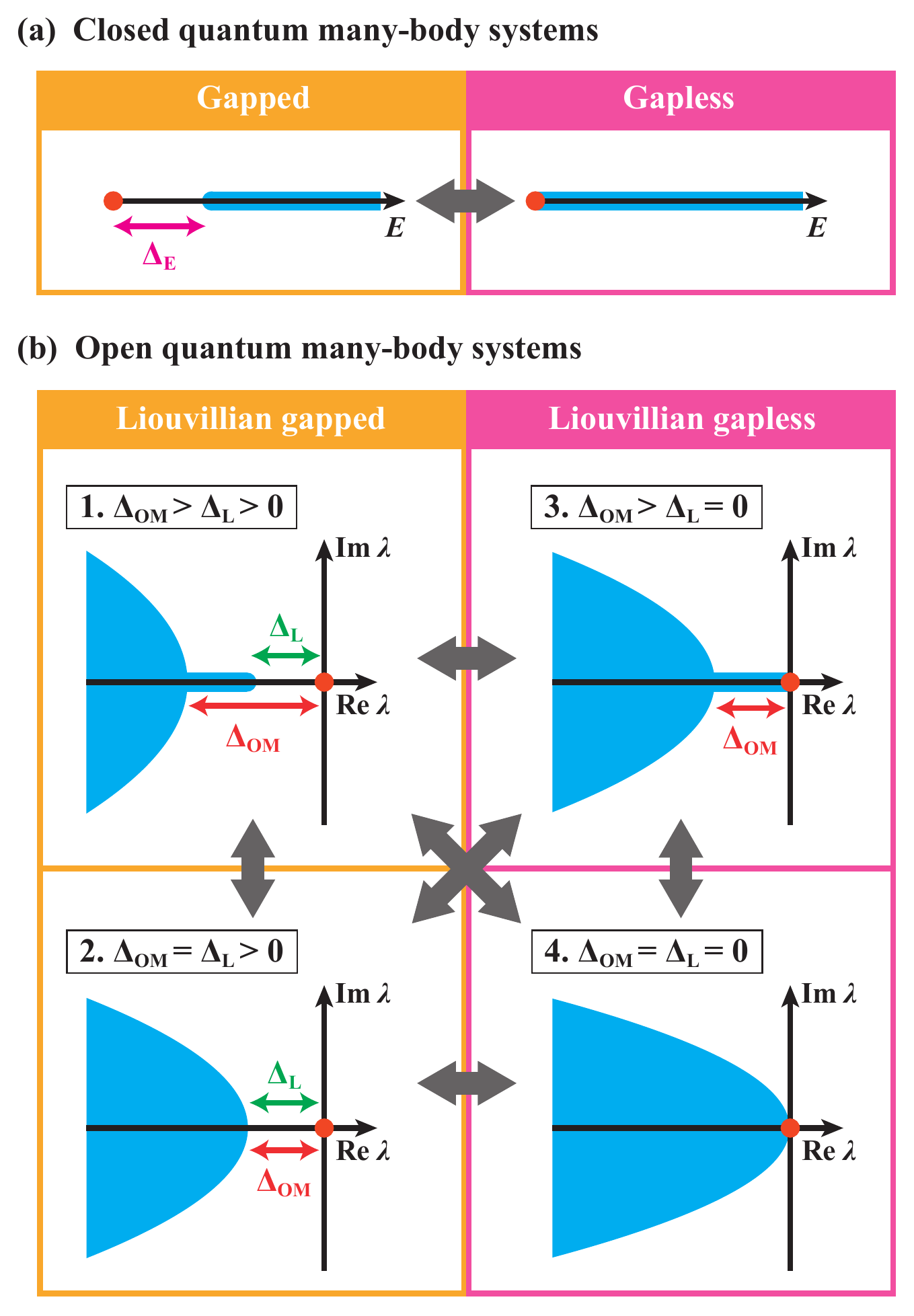}
\caption{Classification of spectral types for (a) closed quantum many-body systems and (b) open quantum many-body systems.
In closed systems (a), spectra of the Hamiltonian are categorized into two types: gapped and gapless, with the red point indicating the ground state. 
Transitions between these states manifest as QPTs.
In open systems (b), Liouvillian spectra are divided into four distinct categories: (1) $\Delta_\mathrm{OM}>\Delta_L>0$, (2) $\Delta_\mathrm{OM}=\Delta_L>0$, (3) $\Delta_\mathrm{OM}>\Delta_L=0$, and (4) $\Delta_\mathrm{OM}=\Delta_L=0$, with the red point indicating the steady state.
The diagram highlights six possible types of DQPTs connecting these categories, indicated by double arrows.}
\label{fig_spectrum_class}
\end{figure}

Initially, we review the characteristics of the Liouvillian gap and its implications for phase transitions within open quantum many-body systems. 
The eigenmodes, denoted as $\rho_\alpha$, and the corresponding eigenvalues $\lambda_\alpha$ of the Liouvillian $\mathcal{L}$ satisfy the relation
\begin{equation}
\mathcal{L}(\rho_\alpha) = \lambda_\alpha \rho_\alpha \quad (\alpha=0, 1, ... , D^2-1),
\label{eigen_eq}
\end{equation}
where $D$ denotes the dimension of the system's Hilbert space.
Among these, the eigenvalue zero is associated with the system's steady states $\rho_0 = \rho_\mathrm{ss}$, and the negative real parts of all other nonzero eigenvalues guarantee the decay of any perturbations to the steady state.
The general properties of the Liouvillian eigenmodes and eigenvalues are summarized in Appendix \ref{appendix_general_properties_of_Liouvillian_eigenmodes}.
The time evolution of the system's density matrix $\rho(t)$ can be written as
\begin{equation}
\rho(t) = \rho_{\mathrm{ss}} + \sum_{\lambda_\alpha \neq 0} c_{\alpha} e^{\lambda_{\alpha} t} \rho_{\alpha},
\label{rho_evolution}
\end{equation}
where $c_{\alpha}$ denotes the coefficients of the eigenmode expansion of the initial density matrix $\rho(0)$.

The Liouvillian gap $\Delta_L$ is formally defined by
\begin{equation}
\Delta_L = \min_{\lambda_\alpha \neq 0} |\mathrm{Re}[\lambda_\alpha]|,
\label{def_Delta_L}
\end{equation}
with $\min_{\lambda_\alpha \neq 0}$ indicating minimization across all nonzero eigenvalues.
Consequently, $\Delta_L$ represents the decay rate of the most slowly-decaying mode.
In numerous examples, the inverse of $\Delta_L$ is identical to the relaxation timescale of the system (however, exceptions to this observation have been documented in Refs.~\cite{Mori-20, Haga-21, Bensa-21, Mori-23}).

A dissipative quantum phase transition (DQPT) is characterized by a singular behavior in the steady state $\rho_\mathrm{ss}$ as a function of a control parameter \cite{Kessler-12, Honing-12, Horstmann-13, Casteels-16, Casteels-17, Fitzpatrick-17, Vicentini-18, Minganti-18, Imamoglu-18, Rota-18, Ferreira-19, Tomadin-11, Lee-11, Torre-13, Lee-13, Ludwig-13, Carr-13, Sieberer-13, Sieberer-14, Marcuzzi-14, Weimer-15, Maghrebi-16, Sieberer-16, Biondi-17, Domokos-17, Young-20}.
The Liouvillian gap $\Delta_L$ serves as a critical indicator for identifying DQPTs \cite{Kessler-12, Honing-12, Horstmann-13, Casteels-16, Casteels-17, Fitzpatrick-17, Vicentini-18, Minganti-18, Imamoglu-18, Rota-18, Ferreira-19}.
Specifically, in scenarios of first-order DQPTs, where $\rho_\mathrm{ss}$ exhibits discontinuous changes, $\Delta_L$ becomes zero exclusively at the transition point, maintaining nonzero values in the distinct phases on either side of the transition \cite{Minganti-18}.
Conversely, during second-order DQPTs, marked by a cuspy behavior of $\rho_\mathrm{ss}$ with respect to the control parameter, $\Delta_L$ vanishes across the entire parameter range within the symmetry-broken phase \cite{Minganti-18}.
Our study concentrates on second-order DQPTs, particularly those involving the spontaneous breaking of a continuous symmetry, such as the $U(1)$ symmetry in bosonic systems.

We define the oscillating-mode (OM) gap as follows:
\begin{equation}
\Delta_\mathrm{OM} = \min_{\mathrm{Im}[\lambda_\alpha] \neq 0} |\mathrm{Re}[\lambda_\alpha]|,
\label{def_Delta_OM}
\end{equation}
with $\min_{\mathrm{Im}[\lambda_\alpha] \neq 0}$ indicating minimization across all eigenvalues possessing a nonzero imaginary part.
This gap represents the decay rate of the most slowly-decaying oscillating mode.
According to this definition, it holds that $\Delta_L \leq \Delta_\mathrm{OM}$.
Note that the concept of OM gap is inapplicable to Hermitian systems, which are characterized by a purely real spectrum.

Let us consider the generic structure of Liouvillian spectra for open quantum many-body systems, where the eigenvalues are anticipated to form a continuous spectrum in the thermodynamic limit.
Utilizing the Liouvillian gap $\Delta_L$ and the OM gap $\Delta_\mathrm{OM}$, we categorize the spectra into four distinct scenarios, as illustrated in Fig.~\ref{fig_spectrum_class}(b):
\begin{enumerate}
\item $\Delta_\mathrm{OM} > \Delta_L > 0$: The spectrum includes a continuum of complex eigenvalues and a line of real eigenvalues, distanced from zero by the Liouvillian gap.
\item $\Delta_\mathrm{OM} = \Delta_L > 0$: The spectrum is comprised solely of a continuum of complex eigenvalues, detached from zero.
\item $\Delta_\mathrm{OM} > \Delta_L = 0$: The spectrum includes a continuum of complex eigenvalues and a line of real eigenvalues extending to zero.
\item $\Delta_\mathrm{OM} = \Delta_L = 0$: The spectrum is comprised solely of a continuum of complex eigenvalues that reach zero.
\end{enumerate}
Spectra Types 1 and 2 are characterized by an open Liouvillian gap, whereas Types 3 and 4 exhibit a closed gap. 
This framework allows for the identification of six possible types of DQPTs that link these spectra.
In the following sections, we demonstrate that dissipative lattice boson models can display all of these spectra types and undergo transitions among them.

The rest of this paper is organized as follows.
In Sec.~\ref{sec_model}, we introduce the dissipative lattice boson models under investigation, focusing on two models distinguished by their unique symmetries. 
These models incorporate dissipation mechanisms that drive the system towards a Bose-Einstein condensed state, augmented with additional dephasing, as well as particle gain and loss mechanisms.
In Sec.~\ref{sec_mean_field_spectra}, we discuss the Liouvillian spectral structure of these models using a mean-field approximation, which enables the analysis of larger system spectra despite neglecting entanglement across different lattice sites.
We demonstrate that our models display the all spectral types introduced in Sec.~\ref{sec_oscillating_mode_gap} and investigate DQPTs among them.
In Sec.~\ref{sec_exact_spectra}, we examine the exact Liouvillian spectra through numerical diagonalization, a method limited by small system size but insightful for observing unique behaviors of the OM gap associated with DQPTs.
In Sec.~\ref{sec_relaxation_dynamics}, we discuss how the closing of the OM gap alters the system's relaxation dynamics toward its steady state.
Finally, Sec.~\ref{sec_conclusions} provides conclusions and perspectives on future research directions.

\section{Model}
\label{sec_model}

We consider a system of interacting bosons on a lattice.
For the one-dimensional case, the Hamiltonian is given by 
\begin{equation}
H = \sum_{j=1}^L \left[ - J  \left( b_j^\dag b_{j+1} + b_{j+1}^\dag b_j \right) + \frac{U}{2} b_j^\dag b_j^\dag b_j b_j - \mu b_j^\dag b_j \right],
\label{Hamiltonian}
\end{equation}
where $b_j$ denotes the boson annihilation operator at site $j$.
Here, $J$ represents the hopping amplitude facilitating boson movement between adjacent sites, $U$ quantifies the on-site interaction strength, and $\mu$ is the chemical potential.
We implement periodic boundary conditions, thereby setting $b_{L+j}=b_j$.
In all numerical calculations, the hopping amplitude $J$ is set to unity.

We consider various dissipation mechanisms, characterized by the following jump operators:
\begin{enumerate}
\item Bond dissipation is modeled by the operator
\begin{equation}
L_{b,j} = \sqrt{\kappa} (b_j^\dag + b_{j+1}^\dag) (b_j - b_{j+1}),
\label{L_bond}
\end{equation}
facilitating the development of phase coherence \cite{Diehl-08}.

\item Dephasing is modeled by the operator
\begin{equation}
L_{d,j} = \sqrt{\gamma} b_j^\dag b_j,
\label{L_dephasing}
\end{equation}
accounting for the loss of coherence without energy dissipation.

\item One-particle pumping is modeled by the operator
\begin{equation}
L_{p,j} = \sqrt{r_p} b_j^\dag,
\label{L_pumping}
\end{equation}
facilitating the addition of particles to the system.

\item  One-particle loss is modeled by the operator
\begin{equation}
L_{l,j} = \sqrt{r_l} b_j,
\label{L_1P_loss}
\end{equation}
corresponding to the removal of particles.

\item Two-particle loss is modeled by the operator
\begin{equation}
L_{t,j} = \sqrt{r_t} b_j b_j,
\label{L_2P_loss}
\end{equation}
describing the simultaneous removal of two particles.
\end{enumerate}

In particular, we focus on two distinct scenarios: 
\begin{enumerate}
\item Model 1: $r_p = r_l = r_t = 0$.
The Liouvillian reads
\begin{align}
\mathcal{L}_1(\rho) =& -i[H, \rho] + \sum_{j=1}^L \left( L_{b,j} \rho L_{b,j}^{\dag} - \frac{1}{2} \{ L_{b,j}^{\dag}L_{b,j}, \rho \} \right) \nonumber \\
&+ \sum_{j=1}^L \left( L_{d,j} \rho L_{d,j}^{\dag} - \frac{1}{2} \{ L_{d,j}^{\dag}L_{d,j}, \rho \} \right).
\label{Liouvillian_model_1}
\end{align}

\item Model 2: $\gamma=0$, $r_t >0$.
The Liouvillian reads
\begin{align}
\mathcal{L}_2(\rho) =& -i[H, \rho] + \sum_{j=1}^L \left( L_{b,j} \rho L_{b,j}^{\dag} - \frac{1}{2} \{ L_{b,j}^{\dag}L_{b,j}, \rho \} \right) \nonumber \\
&+ \sum_{j=1}^L \left( L_{p,j} \rho L_{p,j}^{\dag} - \frac{1}{2} \{ L_{p,j}^{\dag}L_{p,j}, \rho \} \right) \nonumber \\
&+ \sum_{j=1}^L \left( L_{l,j} \rho L_{l,j}^{\dag} - \frac{1}{2} \{ L_{l,j}^{\dag}L_{l,j}, \rho \} \right) \nonumber \\
&+ \sum_{j=1}^L \left( L_{t,j} \rho L_{t,j}^{\dag} - \frac{1}{2} \{ L_{t,j}^{\dag}L_{t,j}, \rho \} \right).
\label{Liouvillian_model_2}
\end{align}
\end{enumerate}
Model 1 is relevant to ultracold atoms on an optical lattice.
The bond dissipation mechanism $L_{b,j}$, aimed at driving the system toward a Bose-Einstein Condensate (BEC) state, can be implemented by immersing ultracold atoms in a BEC bath \cite{Tomadin-11, Diehl-08, Kraus-08, Bonnes-14, Griessner-06}.
Additionally, the dephasing mechanism $L_{d,j}$, which results in heating the system to higher temperatures, can be introduced by intensity fluctuations of the lasers forming the lattice \cite{Pichler-13, Stannigel-14, Cai-13}.
Model 2 represents a discretized version of exciton-polariton systems, which are open interacting bosonic systems maintained by a balance between incoherent pumping and losses of quasiparticles \cite{Deng-10, Carusotto-13, Byrnes-14}.
In this context, the bond dissipation mechanism $L_{b,j}$ serves as an effective diffusion term, generated upon integrating out short-scale fluctuations \cite{Sieberer-13, Sieberer-14, Sieberer-16}.
The microscopic Hamiltonians leading to the dissipation mechanisms given by Eqs.~\eqref{L_bond}-\eqref{L_2P_loss} are discussed in Appendix \ref{appendix_microscopic_Hamiltonian}.

The basic properties of Model 1 are summarized as follows:
\begin{itemize}
\item The particle number $N=\sum_j b_j^\dag b_j$ is conserved, which follows from $[H, N] = [L_{b,j}, N] = [L_{d,j}, N] = 0$.
In other words, Model 1 has the strong $U(1)$ symmetry \cite{Buca-12}.

\item For $U=0$ and $\gamma=0$, the steady state $\rho_\mathrm{ss}$ is given by the BEC state \cite{Diehl-08}: $\rho_\mathrm{ss} = \ket{\Phi_\mathrm{BEC}} \bra{\Phi_\mathrm{BEC}}$ with
\begin{equation}
\ket{\Phi_\mathrm{BEC}} = \frac{1}{\sqrt{N!}}(\tilde{b}_{k=0}^\dag)^N \ket{\mathrm{v}},
\end{equation}
where $\ket{\mathrm{v}}$ is a vacuum state, and $\tilde{b}_k$ is the annihilation operator for a wavenumber $k=2\pi n / L \: (n = -L/2+1,...,L/2)$,
\begin{equation}
\tilde{b}_k = \frac{1}{\sqrt{L}} \sum_{j=1}^L e^{ikj} b_j.
\end{equation}
The equation $\mathcal{L}(\ket{\Phi_\mathrm{BEC}} \bra{\Phi_\mathrm{BEC}})=0$ follows from the fact that $\ket{\Phi_\mathrm{BEC}}$ is an eigenstate of $H$ with $U=0$ and
\begin{equation}
(b_j - b_{j+1}) \ket{\Phi_\mathrm{BEC}} = \frac{1}{\sqrt{L}} \sum_k (1-e^{-ik}) e^{-ikj} \tilde{b}_k \ket{\Phi_\mathrm{BEC}} = 0.
\end{equation}

\item As discussed in Sec.~\ref{sec_mean_field_spectra_1}, within the mean-field approximation, the steady state exhibits off-diagonal long-range order, ``superfluid", for small values of $U$ and $\gamma$.
When $U$ or $\gamma$ is increased, a second-order phase transition to a disordered phase, ``normal fluid", occurs. 

\item As discussed in Sec.~\ref{sec_mean_field_spectra_1}, for both ordered and disordered phases, the Liouvillian gap $\Delta_L$ vanishes in the thermodynamic limit.
Specifically, $\Delta_L$ scales as $L^{-2}$ with respect to the system size $L$.
\end{itemize}

The basic properties of Model 2 are summarized as follows:
\begin{itemize}
\item The particle number $N=\sum_j b_j^\dag b_j$ is not conserved due to the presence of both pumping and loss mechanisms.
However, the Liouvillian \eqref{Liouvillian_model_2} is invariant with respect to the transformation $b_j \to e^{i\theta} b_j$, $b_j^\dag \to e^{-i\theta} b_j^\dag$.
In other words, Model 2 has the weak $U(1)$ symmetry \cite{Buca-12}.

\item The particle density is determined by the balance between the pumping rate $r_p$ and the loss rate $r_l$.
The two-particle loss is included to preclude the unbounded increase of the particle number.

\item As discussed in Sec.~\ref{sec_mean_field_spectra_2}, within the mean-field approximation, the steady state exhibits off-diagonal long-range order,  ``superfluid", for sufficiently large $r_p$.
An increase in the interaction strength $U$ triggers a second-order phase transition, leading to a disordered phase,  or ``normal fluid".

\item As discussed in Sec.~\ref{sec_mean_field_spectra_2}, the Liouvillian gap $\Delta_L$ diminishes to zero within the ordered phase, whereas it becomes finite in the disordered phase, highlighting distinct dynamical properties across these phases.
\end{itemize}

While both models facilitate the emergence of a $U(1)$ symmetry-broken phase within their steady states, their spectral structures exhibit qualitatively distinctive characteristics, reflecting different symmetries.
In subsequent sections, we will discuss the classification of the spectral structures of these models according to the categories illustrated in Fig.~\ref{fig_spectrum_class}.

\section{Mean-field spectra}
\label{sec_mean_field_spectra}

To elucidate the connection between phase structure and spectral properties, we employ the mean-field approximation, wherein the many-body density matrix $\rho$ is approximated as a product of single-site states, $\rho \approx \tilde{\rho}_1 \otimes \tilde{\rho}_2 \otimes \cdots \otimes \tilde{\rho}_L$, with $\tilde{\rho}_j$ representing the single-site density matrix at site $j$.
The mean-field master equation for $\tilde{\rho}_j$ is expressed as
\begin{equation}
\partial_t \tilde{\rho}_j = \mathcal{L}_\mathrm{MF}( \tilde{\rho}_j; \tilde{\rho}_{j-1}, \tilde{\rho}_{j+1})  \quad (j = 1, ..., L),
\label{master_eq_mf}
\end{equation}
where the right-hand side is a quadratic function of the matrix elements of $\tilde{\rho}_j$, $\tilde{\rho}_{j-1}$, and $\tilde{\rho}_{j+1}$, with its detailed form presented in Appendix \ref{appendix_mean_field_master_equation}.
By solving Eq.~\eqref{master_eq_mf} for sufficiently long time, $\tilde{\rho}_j$ is expected to approach a steady state $\tilde{\rho}_{\mathrm{ss},j}$.
It is known that non-uniform steady states with periodic density modulation emerge for $J \ll \kappa$ and $\gamma=0$ \cite{Tomadin-11}.
However, this study concentrates on a regime characterized by uniform steady states.
Consequently, we simplify notation by excluding the subscript $j$ in $\tilde{\rho}_{\mathrm{ss},j}$.
The superfluid order parameter is defined as
\begin{equation}
\psi := \langle b_j \rangle = \mathrm{tr}\left[ \tilde{\rho}_\mathrm{ss} b_j \right].
\end{equation}
and the condensate density is given by $n_0=|\psi|^2$.
A nonzero value of $n_0$ signifies the superfluid phase, whereas its absence indicates a normal fluid phase.

To obtain the steady state, it is necessary to adjust the chemical potential $\mu$ to remove the phase rotation of $\psi$.
The determination of $\mu$ proceeds as follows.
Initially, the mean-field master equation \eqref{master_eq_mf} is numerically solved starting with $\mu=0$, and the time-dependent order parameter $\psi_j(t)=\mathrm{tr}[ \tilde{\rho}_j(t) b_j ]$ is calculated.
Subsequently, the phase rotation frequency of $\psi_j(t)$ is derived via
\begin{equation}
\omega = -i \lim_{t \to \infty} \frac{\partial_t \psi_j(t)}{\psi_j(t)}.
\label{omega_psi}
\end{equation}
In the superfluid phase, $\omega$ obtained from Eq.~\eqref{omega_psi} is real.
By assigning $\mu=-\omega$ in Eq.~\eqref{master_eq_mf}, we have a time-independent state after a long time.
Conversely, in the normal fluid phase, as $\psi_j(t)$ tends towards zero, we have $\mathrm{Im}[\omega] >0$.
Herein, we set $\mu=-\mathrm{Re}[\omega]$.

The excitation spectrum is obtained by linearizing the mean-field master equation \eqref{master_eq_mf} with respect to a small deviation from the steady state, $\delta \tilde{\rho}_j = \tilde{\rho}_j - \tilde{\rho}_\mathrm{ss}$.
This linearization yields the equation:
\begin{equation}
\partial_t \delta \tilde{\rho}_j = \mathcal{M} ( \delta \tilde{\rho}_j; \delta \tilde{\rho}_{j-1}, \delta \tilde{\rho}_{j+1})  \quad (j = 1, ..., L),
\label{master_eq_mf_linearized}
\end{equation}
where the right-hand side is a linear function of the matrix elements of $\tilde{\rho}_j$, $\tilde{\rho}_{j-1}$, and $\tilde{\rho}_{j+1}$, with its detailed form presented in Appendix \ref{appendix_mean_field_master_equation}.
The corresponding eigenmodes $\tilde{\rho}_j^\alpha$ and eigenvalues $\lambda^\mathrm{MF}_\alpha$ satisfy
\begin{equation}
\mathcal{M}( \tilde{\rho}_j^\alpha; \tilde{\rho}_{j-1}^\alpha, \tilde{\rho}_{j+1}^\alpha) = \lambda^\mathrm{MF}_\alpha \tilde{\rho}_j^\alpha \quad (j = 1, ..., L).
\label{eigenvalue_L_mf}
\end{equation}
Note that $\lambda^\mathrm{MF}_\alpha$ is independent of the site index $j$.
The collection of $\lambda^\mathrm{MF}_\alpha$ constitutes the mean-field spectrum.
It is important to recall that the eigenvalues of the Liouvillian appear as complex conjugate pairs, ensuring that the spectrum is symmetric with respect to the real axis (see Appendix \ref{appendix_general_properties_of_Liouvillian_eigenmodes}).
This property holds true not only for the exact eigenvalues of the Liouvillian but also for the mean-field eigenvalues $\lambda^\mathrm{MF}_\alpha$.

The relationship between the exact Liouvillian spectrum and its mean-field spectrum is nontrivial.
The exact Liouvillian spectrum includes
\begin{equation}
D^2 = \binom{L+N-1}{N}^2
\end{equation}
eigenvalues, where $D$ is the dimension of the Hilbert space of the system.
The dimension $D$ corresponds to the number of ways to distribute $N$ indistinguishable particles across $L$ sites, which is $D=\binom{L+N-1}{N}$ from the ``stars and bars" argument.
On the other hand, the mean-field spectrum consists of $d_\mathrm{max}^2 L$ eigenvalues, where $d_\mathrm{max}$ is the cutoff dimension of the single-site Hilbert space.
By neglecting inter-site entanglement, the mean-field approximation omits entangled eigenmodes from the Liouvillian spectrum.
Nonetheless, in systems of high spatial dimensions, the linearized master equation \eqref{master_eq_mf_linearized} accurately approximates the dynamical evolution of excitations around the steady state.
This observation implies that the mean-field spectrum can serve as a reasonable surrogate for the exact spectrum near the steady state.
Specifically, the gaps $\Delta_L^\mathrm{MF}$ and $\Delta_\mathrm{OM}^\mathrm{MF}$, derived from the mean-field spectrum, are anticipated to converge to their exact counterparts $\Delta_L$ and $\Delta_\mathrm{OM}$ in the limit of infinite spatial dimensions.

We mention the numerical procedures adopted for calculating the mean-field spectra.
To obtain the steady state, we solve the mean-field master equation \eqref{master_eq_mf} by employing a fourth-order Runge-Kutta integration method with a time step of $dt = 0.005$.
The cutoff dimension $d_\mathrm{max}$ of the single-site Hilbert space is set to $20$, beyond which no significant changes in the mean-field spectra near the steady state are observed.
To enhance computational efficiency in the diagonalization of the mean-field Liouvillian $\mathcal{M}$, we exploit its translational invariance. 
Specifically, we employ the following ansatz for the eigenmodes \cite{Tomadin-11}:
\begin{equation}
\tilde{\rho}_j^\alpha \propto e^{i \phi j} \quad (\phi=2\pi m/L, \: m=0,1,...,L-1),
\label{uniform_eigenmode_assumption}
\end{equation}
facilitating the reduction of the eigenvalue problem described by Eq.~\eqref{eigenvalue_L_mf} to a single-site problem.
We confirm that this simplified approach yields results identical to those obtained through the direct diagonalization of the mean-field Liouvillian $\mathcal{M}$.

\subsection{Model 1}
\label{sec_mean_field_spectra_1}

\begin{figure}
\centering
\includegraphics[width=0.45\textwidth]{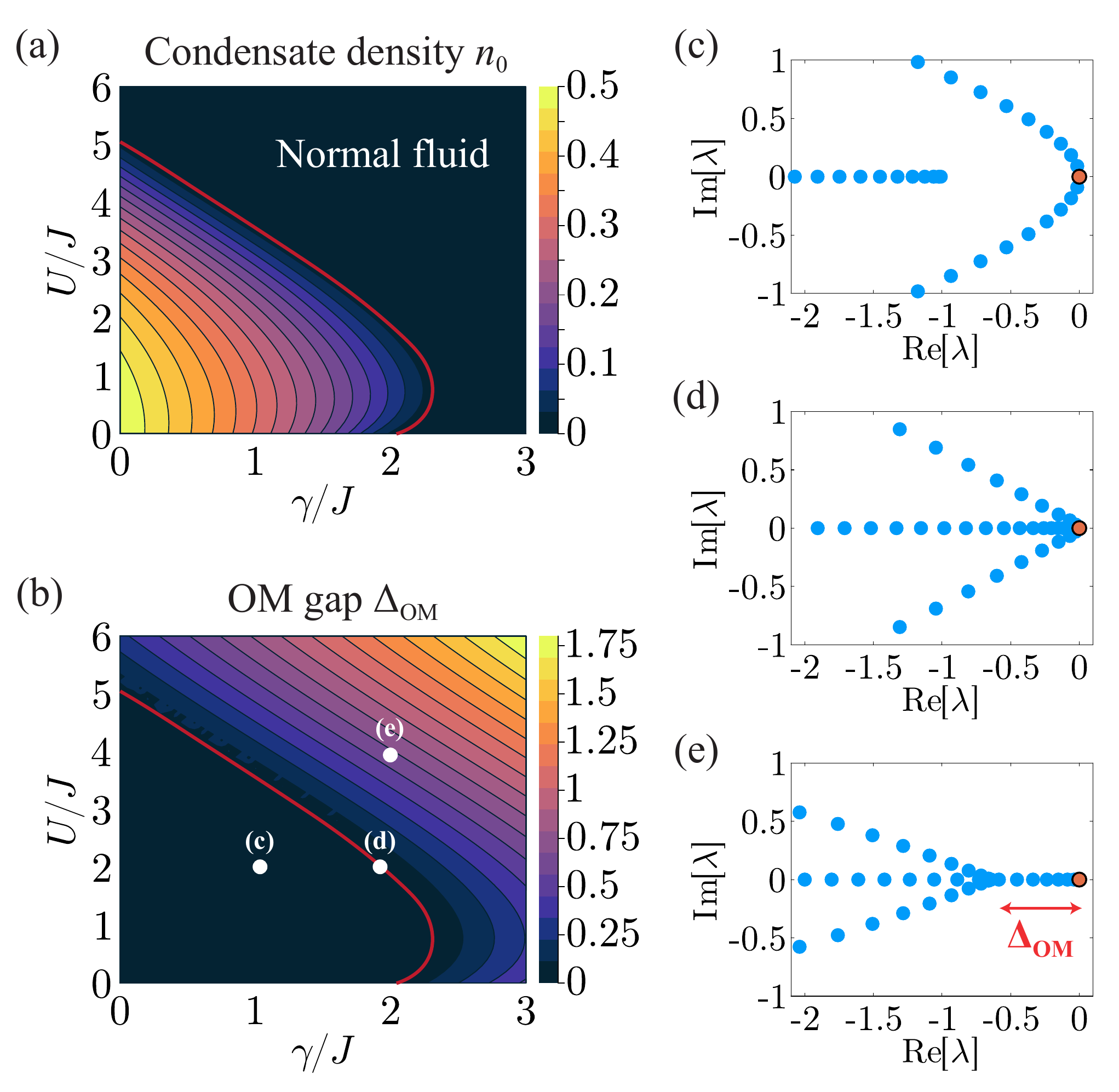}
\caption{OM gap and mean-field spectra for Model 1.
(a) Condensate density $n_0 = |\psi|^2$ as a function of $\gamma/J$ and $U/J$.
The bright region indicates the superfluid phase ($n_0 > 0$) and the dark region denotes the normal fluid phase ($n_0 = 0$).
(b) OM gap $\Delta_\mathrm{OM}$ as a function of $\gamma/J$ and $U/J$.
The red solid curves in (a) and (b) represent the phase boundary determined from $n_0$.
(c)-(e) Mean-field spectra for (1) $\gamma/J=1$, $U/J=2$, (2) $\gamma/J=1.9$, $U/J=2$, and (3) $\gamma/J=2$, $U/J=4$, highlighted by white dots in (b).
The particle density is $N/L=0.5$, with a system size $L=64$ and $\kappa=1$.
The transition from the normal fluid to superfluid is characterized by the closing of the OM gap.}
\label{fig_model1_OM_gap}
\end{figure}

For Model 1, Fig.~\ref{fig_model1_OM_gap}(a) presents the condensate density $n_0=|\psi|^2$ for the steady state, mapped as a function of the dephasing rate $\gamma/J$ and the interaction strength $U/J$.
Here, bright and dark regions represent the superfluid and normal fluid phases, respectively. 
Note that, at $\gamma=U=0$, $n_0$ is equal to the particle density $n=N/L$ because the single-site steady state $\tilde{\rho}_\mathrm{ss}$ is given by the coherent state, $\tilde{\rho}_\mathrm{ss} = \ket{\psi}\bra{\psi}$ with
\begin{equation}
\ket{\psi} = \sum_{\nu=0}^\infty e^{-|\psi|^2/2} \frac{\psi^\nu}{\sqrt{\nu !}} \ket{\nu},
\label{coherent_state}
\end{equation} 
where $\ket{\nu}$ denotes the basis state with $\nu$ particles.
With increases in $\gamma$ and $U$, $n_0$ diminishes, leading to a phase transition to the normal fluid phase.
Intriguingly, the phase boundary exhibits a reentrant behavior, indicating that the critical dephasing rate $\gamma_c$ varies non-monotonically with $U$.

Figure \ref{fig_model1_OM_gap}(b) displays the OM gap $\Delta_\mathrm{OM}^\mathrm{MF}$ obtained from the mean-field spectra, where the phase boundary is indicated by the red solid curve.
It is noted that $\Delta_\mathrm{OM}^\mathrm{MF}=0$ in the superfluid phase, whereas $\Delta_\mathrm{OM}^\mathrm{MF}>0$ in the normal fluid phase.
Figures \ref{fig_model1_OM_gap}(c)-(e) illustrate typical mean-field spectra at the superfluid phase, phase boundary, and normal fluid phase, respectively.
For both phases, the Liouvillian gap $\Delta_L^\mathrm{MF}$ vanishes, scaling as $L^{-2}$ with the system size $L$.
In the superfluid phase , the spectrum features a line of real eigenvalues and complex eigenvalues forming a parabolic curve [see Fig.~\ref{fig_model1_OM_gap}(c)].
At the critical point, the line of real eigenvalues touches the origin [see Fig.~\ref{fig_model1_OM_gap}(d)].
In the normal fluid phase, the OM gap opens due to a shift of complex eigenvalues towards the negative real axis [see Fig.~\ref{fig_model1_OM_gap}(e)].
In summary, the phase transition in Model 1, marked by the closure of the OM gap, coincides with a transition from Type 3 to Type 4 spectra, depicted in Fig.~\ref{fig_spectrum_class}.

We present a theoretical analysis to explain the emergence of the OM gap for large dephasing rates within the context of Model 1.
Assuming the system is in the normal fluid phase with a large $\gamma$, the intersite coupling terms in Eq.~\eqref{master_eq_mf} vanishes, yielding a set of decoupled master equations for each site.
The corresponding steady state is then given as \cite{Tomadin-11}
\begin{equation}
\tilde{\rho}_\mathrm{ss} = \sum_{\nu = 0}^\infty \frac{\bar{n}^\nu}{(\bar{n}+1)^\nu} \ket{\nu} \bra{\nu},
\label{rho_ss_normal}
\end{equation}
where $\ket{\nu}$ denotes the basis state with $\nu$ particles, and $\bar{n}=N/L$ is the average particle density.
It should be noted that, in the normal fluid phase, $\tilde{\rho}_\mathrm{ss}$ is independent of other parameters.
Given the diagonal nature of $\tilde{\rho}_\mathrm{ss}$, the linearized master equation for $\delta \tilde{\rho}_j = \tilde{\rho}_j - \tilde{\rho}_\mathrm{ss}$ simplifies significantly.
\begin{widetext}
Representing $\delta \tilde{\rho}_j = \sum_{m, n = 0}^\infty Q_{m,n}^j \ket{m} \bra{n}$, the time evolution of the matrix elements $Q_{m,n}^j$ follows
\begin{equation}
\partial_t Q_{m,n}^j = F^H_{m,n}(Q^j; Q^{j-1}, Q^{j+1}) +  F^b_{m,n}(Q^j; Q^{j-1}, Q^{j+1}) +  F^d_{m,n}(Q^j),
\end{equation}
where $F^H_{m,n}$, $F^b_{m,n}$, and $F^d_{m,n}$ denote contributions from the unitary dynamics, bond dissipation, and dephasing, respectively.
Specifically, the contribution from the unitary dynamics is given by
\begin{align}
F^H_{m,n}(Q^j; Q^{j-1}, Q^{j+1}) =& - i \left[ \frac{U}{2} m(m-1) - \frac{U}{2} n(n-1) - \mu (m - n) \right] Q_{m,n}^j \nonumber \\
&+ i J (p_n^\mathrm{ss} - p_m^\mathrm{ss}) \left[ \sqrt{m} (\psi_{j+1} + \psi_{j-1}) \delta_{m,n+1} + \sqrt{n} (\bar{\psi}_{j+1} + \bar{\psi}_{j-1}) \delta_{m+1,n} \right],
\label{d_Q_H}
\end{align}
where $p_\nu^\mathrm{ss} = \bar{n}^\nu / (\bar{n}+1)^\nu$ denotes the diagonal elements of $\tilde{\rho}_\mathrm{ss}$, and
$\psi_j = \sum_{\nu=0}^\infty \sqrt{\nu+1} Q_{\nu+1, \nu}^j$ and $\bar{\psi}_j = \sum_{\nu=0}^\infty \sqrt{\nu+1} Q_{\nu, \nu+1}^j$.
The contribution from dephasing reads
\begin{equation}
F^d_{m,n}(Q^j) = - \frac{\gamma}{2} (m-n)^2 Q_{m,n}^j.
\label{d_Q_d}
\end{equation}
The explicit form of $F^b_{m,n}$ is omitted here due to its complexity, however, focusing on its diagonal component, we find
\begin{align}
F^b_{n,n}(Q^j; Q^{j-1}, Q^{j+1}) =& \ 2 \kappa \bar{n} \left[ n Q_{n-1, n-1}^j - (n+1) Q_{n,n}^j \right] + 2 \kappa (\bar{n}+1) \left[ (n+1) Q_{n+1, n+1}^j - n Q_{n,n}^j \right] \nonumber \\
&+ \kappa \left[ n p_{n-1}^\mathrm{ss} - (2n+1) p_n^\mathrm{ss} + (n+1) p_{n+1}^\mathrm{ss} \right] ( \langle \delta n_{j+1} \rangle + \langle \delta n_{j-1} \rangle),
\label{d_Q_b}
\end{align}
where $\langle \delta n_j \rangle = \sum_{\nu = 0}^\infty \nu Q_{\nu,\nu}^j$ represents the density perturbation.
\end{widetext}

\begin{figure}[b]
\centering
\includegraphics[width=0.45\textwidth]{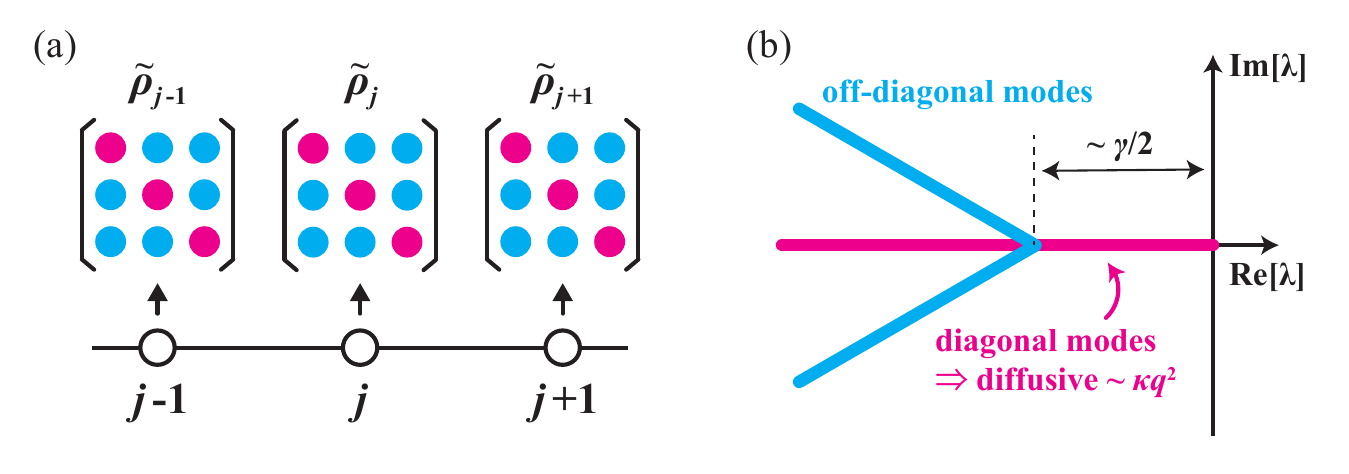}
\caption{(a) Schematic representation of the single-site density matrix $\tilde{\rho}_j$ within the mean-field approximation, where red and blue circles indicate the diagonal and off-diagonal matrix elements, respectively.
(b) Schematic illustration of the mean-field spectrum under conditions of large $\gamma$. 
The diagonal elements of $\tilde{\rho}_j$ correspond to diffusive modes characterized by real eigenvalues $-\kappa q^2$, with $q$ representing the wavenumber associated with the density perturbation. 
In contrast, the off-diagonal elements of $\tilde{\rho}_j$ yield complex eigenvalues, each with a decay rate larger than $\gamma/2$.}
\label{fig_model1_spec_schematic}
\end{figure}

From Eqs.~\eqref{d_Q_H}, \eqref{d_Q_d}, and \eqref{d_Q_b}, we can deduce the spectral structure under conditions of large $\gamma$.
Notably, the unitary dynamics and dephasing do not affect the time evolution of the diagonal elements of the density matrix, as $F^H_{n,n}(Q^j; Q^{j-1}, Q^{j+1})=0$ and $F^d_{n,n}(Q^j)=0$.
According to Eq.~\eqref{d_Q_b}, the time evolution of the density perturbation $\langle \delta n_j \rangle = \sum_{\nu = 0}^\infty \nu Q_{\nu,\nu}^j$ follows
\begin{equation}
\partial_t \langle \delta n_j \rangle = \kappa \left( \langle \delta n_{j+1} \rangle - 2\langle \delta n_j \rangle + \langle \delta n_{j-1} \rangle \right),
\end{equation}
indicating a diffusive relaxation mechanism.
Thus, the Liouvillian eigenvalues associated with density relaxation are real and take the form $-\kappa q^2$, where $q$ represents the wavenumber of the density modulation.
This implies that the Liouvillian gap $\Delta_L$ diminishes as $L^{-2}$.
Conversely, Eq.~\eqref{d_Q_d} implies that the off-diagonal elements decay with rates larger than $\gamma/2$.
Figure \ref{fig_model1_spec_schematic}(b) schematically represents the mean-field spectrum for sufficiently large $\gamma$.
Since the relaxation of the diagonal elements is diffusive, the corresponding eigenvalues are located on the real axis, which are depicted by the red line.
On the other hand, the eigenmodes associated with off-diagonal elements have complex eigenvalues with real parts less than $-\gamma/2$, which are depicted by the blue lines.
This picture elucidates how an increase in $\gamma$ facilitates the opening of the OM gap.

\subsection{Model 2}
\label{sec_mean_field_spectra_2}

\begin{figure}
\centering
\includegraphics[width=0.45\textwidth]{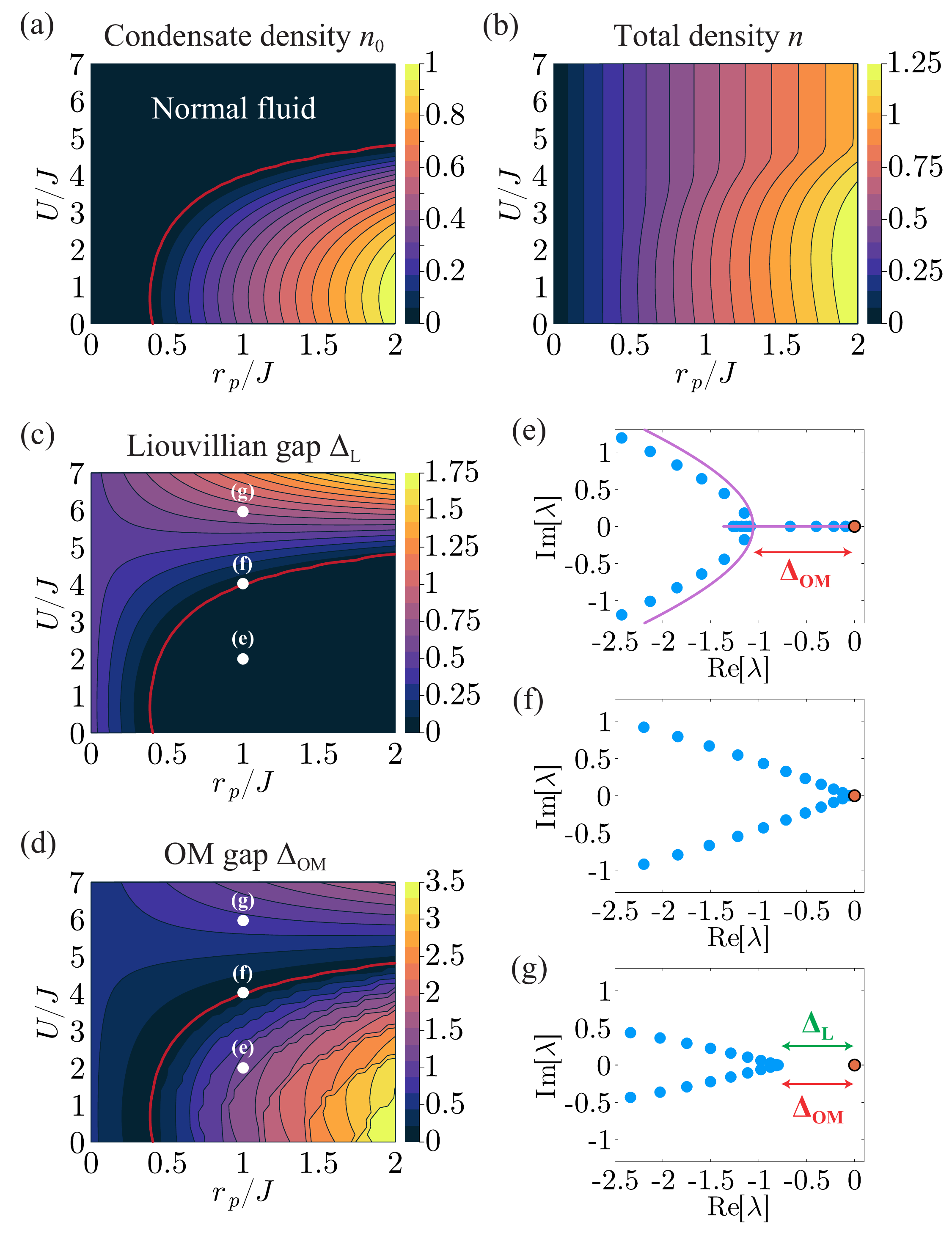}
\caption{OM gap and mean-field spectra for Model 2.
(a) Condensate density $n_0 = |\psi|^2$ as a function of $r_p/J$ and $U/J$.
The bright region indicates the superfluid phase ($n_0 > 0$) and the dark region denotes the normal fluid phase ($n_0 = 0$).
(b) Total density $n$ as a function of $r_p/J$ and $U/J$.
(c) Liouvillian gap $\Delta_L$ as a function of $r_p/J$ and $U/J$.
(d) OM gap $\Delta_\mathrm{OM}$ as a function of $r_p/J$ and $U/J$.
The red solid curves in (a), (c), and (d) represent the phase boundary determined from $n_0$.
(e)-(g) Mean-field spectra for (e) $r_p/J=1$, $U/J=2$, (f) $r_p/J=1$, $U/J=4$, and (g) $r_p/J=1$, $U/J=6$, highlighted by white dots in (c) and (d).
The solid curves in (e) represent the spectrum \eqref{Model2_GP_spectrum} obtained from the GP equation.
The system size is $L=64$, and $\kappa=r_l=r_t=1$.
The transition from the normal fluid to superfluid is characterized by the closing of the Liouvillian gap.}
\label{fig_model2_OM_gap}
\end{figure}

For Model 2, where the particle number is not conserved, we define the total particle density $n$ as follows:
\begin{equation}
n = \mathrm{tr}\left[ \tilde{\rho}_\mathrm{ss} b_j^\dag b_j \right],
\end{equation}
which is expressed as the sum of the condensate density $n_0$ and the noncondensate density $n_1$.
Figures \ref{fig_model2_OM_gap}(a) and (b) show the condensate density $n_0$ and the total density $n$, respectively.
Note that $n$ remains positive for any nonzero pumping rate $r_p$.
In the absence of pumping ($r_p=0$), the steady state is given by the vacuum state, resulting in $n=0$. 
We identify a region with $n_0>0$ as the superfluid phase.
In regimes of high interaction strength $U$, a distinct region emerges where $n_0=0$ and $n>0$, indicative of the normal fluid phase.
Notably, as depicted in Fig.~\ref{fig_model2_OM_gap}(b), $n$ appears to be independent of $U$ in the normal fluid phase, suggesting that the noncondensate particle density is unaffected by interaction-induced depletion.

The Liouvillian gap $\Delta_L$ and the OM gap $\Delta_\mathrm{OM}$ are depicted in Figs.~\ref{fig_model2_OM_gap}(c) and (d), respectively, with the red solid curves marking the phase boundaries determined from $n_0$.
The Liouvillian gap closes in the superfluid phase and opens in the normal fluid phase.
In contrast, the OM gap closes exclusively at the phase boundary and remains open in both the superfluid and normal fluid phases.
The non-smooth contour curves of the OM gap in the superfluid phase are attributable to the presence of exceptional points.
Specifically, deeper within the superfluid phase, the OM gap opens due to the collisions of complex eigenvalue pairs, as suggested in Fig.~\ref{fig_model2_OM_gap}(e).
At these exceptional points, the OM gap exhibits discontinuous jumps.
As the system size approaches infinity, the magnitude of these jumps diminishes, leading to the restoration of smooth contour curves.

Figures \ref{fig_model2_OM_gap}(e)-(g) display typical spectra for the parameters indicated by white dots in (c) and (d).
In the superfluid phase, the spectrum features a line of real eigenvalues and complex eigenvalues forming a parabolic curve [see Fig.~\ref{fig_model2_OM_gap}(e)].
Unlike in Model 1, the OM gap opens here, and the underlying mechanism for this phenomenon will be discussed subsequently.
At the critical point, the curve of complex eigenvalues touches the origin, resulting in the closure of the OM gap [see Fig.~\ref{fig_model2_OM_gap}(f)].
In the normal fluid phase, this curve separates from the origin, and both $\Delta_L$ and $\Delta_\mathrm{OM}$ are greater than zero [see Fig.~\ref{fig_model2_OM_gap}(g)].
These observations indicate that the transition in Model 2 corresponds to a transition from Type 2 to Type 3 spectra, according to the classification in Fig.~\ref{fig_spectrum_class}.

We now present a theoretical analysis that explains the mechanism behind the opening of the OM gap in the superfluid phase of Model 2.
Initially, let us derive the Gross-Pitaevskii (GP) equation for the order parameter $\psi$.
The single-site coherent state is defined by Eq.~\eqref{coherent_state}.
In the deep superfluid phase, where the noncondensate density $n_1 = n - n_0$ is significantly less than the total density $n$, the density matrix is well-approximated by the coherent state $\tilde{\rho}_j = \ket{\psi_j}\bra{\psi_j}$.
The time evolution of $\psi_j$ is given by
\begin{equation}
\partial_t \psi_j = \mathrm{tr}[ b_j \partial_t \tilde{\rho}_j].
\label{GP_Model2_pre}
\end{equation}
Incorporating the mean-field master equation \eqref{master_eq_mf} into the above equation results in the GP equation:
\begin{align}
i \partial_t \psi_j = &- J (\psi_{j+1} + \psi_{j-1}) + U |\psi_j|^2 \psi_j - \mu \psi_j \nonumber \\
&+ i \kappa (\psi_{j+1} - 2\psi_j + \psi_{j-1}) \nonumber \\
&+ i \kappa (|\psi_{j+1}|^2 \psi_{j+1} + |\psi_{j-1}|^2 \psi_{j-1}) \nonumber \\
&- i \kappa (\psi_j^2 \psi_{j+1}^* + \psi_j^2 \psi_{j-1}^*) \nonumber \\
&+ i r_d \psi_j - i r_t |\psi_j|^2 \psi_j,
\label{Model2_GP_eq}
\end{align}
where $r_d = (r_p - r_l)/2$.
The first line represents the conventional GP equation for a lattice.
The subsequent lines account for the contributions from bond dissipation, which drive the system towards uniformity, while the final line represents the effects of pumping and loss.
For a uniform state $\psi_j = \psi_0$, $\psi_0$ vanishes for $r_d < 0$ and $|\psi_0| = \sqrt{r_d/r_t}$ for $r_d > 0$. 
To achieve a time-independent state, the chemical potential $\mu$ must be adjusted to $\mu = -2J + Un_0$, where $n_0 = |\psi_0|^2$.

Linearizing Eq.~\eqref{Model2_GP_eq} with respect to small perturbations $\delta \psi_j = \psi_j - \psi_0$ yields
\begin{align}
i \partial_t \delta \psi_j = &[-J + i \kappa (1+2n_0)] (\delta \psi_{j+1} - 2 \delta \psi_j + \delta \psi_{j-1}) \nonumber \\
&+ (U - i r_t) n_0 (\delta \psi_j + \delta \psi_j^*).
\end{align}
Substituting $\delta \psi_j = u_k e^{i k j + \lambda t} + v_k e^{-i k j + \lambda^* t}$, the decay rate of excitation modes with a wavenumber $k$ is calculated as
\begin{align}
\lambda = &- 2 \kappa (1+2n_0) (1 - \cos k) - r_t n_0 \nonumber \\
&\pm i \sqrt{[2J(1 - \cos k)+Un_0]^2 - (U^2 + r_t^2) n_0^2}.
\label{Model2_GP_spectrum}
\end{align}
When $\kappa=0$ and $r_t=0$, Eq.~\eqref{Model2_GP_spectrum} simplifies to the standard Bogoliubov result, where for $k \to 0$ the dispersion of excitation modes is phononic, $\lambda = \pm i \sqrt{2JUn_0} k$.
With $\kappa$ and $r_t$ present, the dispersion characteristics are substantially altered.
For $k \to 0$, Eq.~\eqref{Model2_GP_spectrum} becomes
\begin{align}
\lambda \simeq - \left[ \kappa (1+2n_0) + \frac{JU}{r_t} \right] k^2, \quad -2r_t n_0.
\label{Model2_GP_spectrum_small_k}
\end{align}
This equation suggests that the spectrum includes a line of real eigenvalues linked to the origin, indicative of a nonzero OM gap.
The occurrence of purely diffusive, non-propagating modes is attributed to the dissipative Goldstone mode, emerging from the spontaneous $U(1)$ symmetry breaking in the superfluid phase \cite{Wouters-06, Wouters-07, Szymanska-06}.
In Fig.~\ref{fig_model2_OM_gap}(e), the spectrum given by Eq.~\eqref{Model2_GP_spectrum} is depicted by solid lines, where the condensate density $n_0$ in Eq.~\eqref{Model2_GP_spectrum} is replaced with the total density $n$ to achieve a good fit.
At the transition point, the spectrum can be obtained by setting $n_0=0$ in Eq.~\eqref{Model2_GP_spectrum}:
\begin{equation}
\lambda = - 2 \kappa (1 - \cos k) \pm i 2J(1 - \cos k).
\end{equation}
Thus, the spectrum is linear near the origin, $|\mathrm{Re}[\lambda]|/|\mathrm{Im}[\lambda]| = \kappa/J$, which is consistent with Fig.~\ref{fig_model2_OM_gap}(f).

\begin{figure}
\centering
\includegraphics[width=0.45\textwidth]{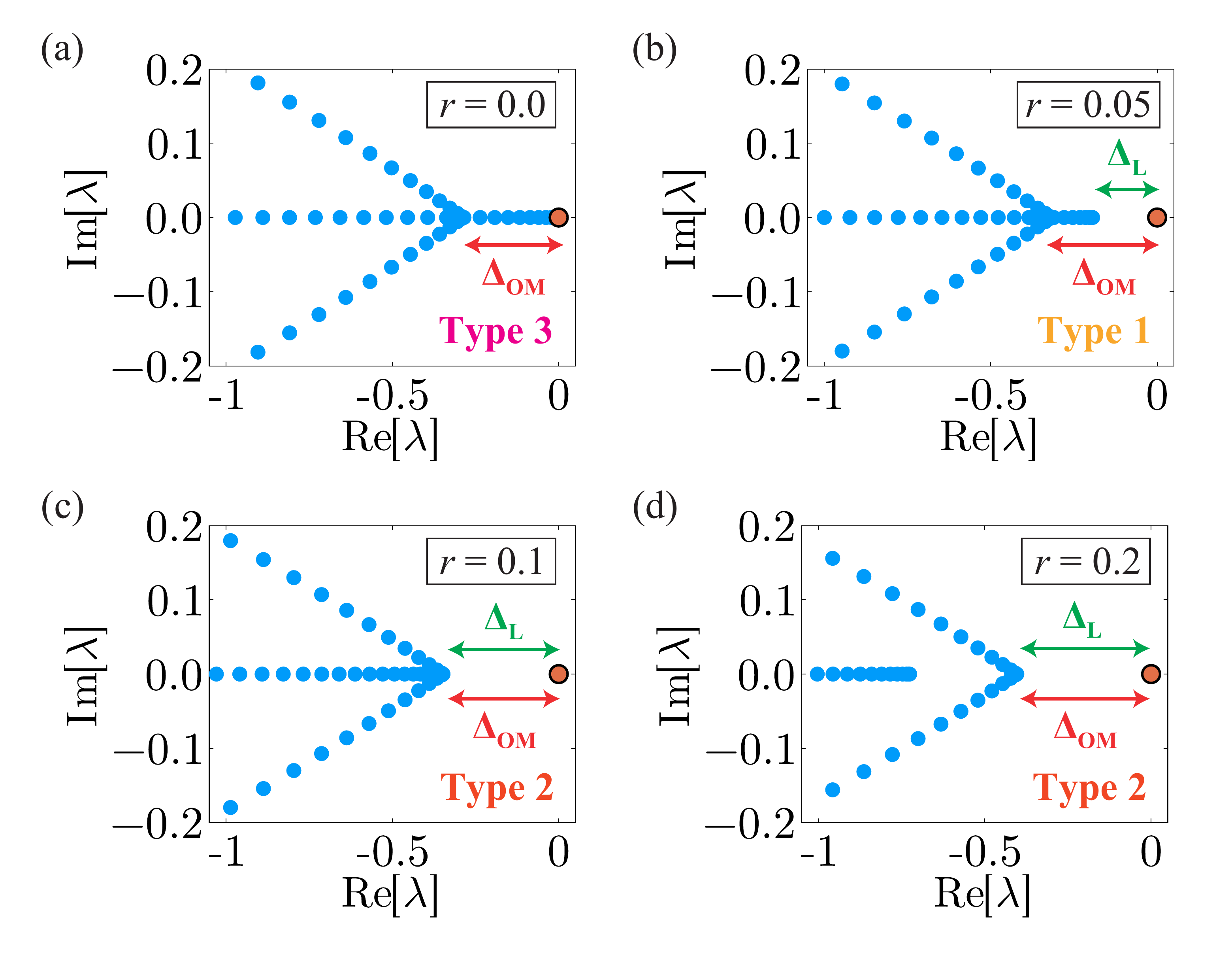}
\caption{Spectral transition from Model 1 to Model 2 within the normal fluid phase.
The pumping and loss rates are uniformly adjusted, $r := r_p=r_l=r_t$, across panels (a) $r=0$, (b) $r=0.05$, (c) $r=0.1$, and (d) $r=0.2$.
The system size is $L=128$, with a dephasing rate $\gamma=0$ and interaction strength $U=6$.
Increasing $r$ induces successive transitions among Type 3, Type 1, and Type 2 spectra.}
\label{fig_model2_lambda_small_r}
\end{figure}

We examine the transition from Model 1 to Model 2.
Figure \ref{fig_model2_lambda_small_r} displays the evolution of mean-field spectra as the unified rate $r := r_p=r_l=r_t$ increases within the normal fluid phase.
The scenario with $r=0$, depicted in Fig.~\ref{fig_model2_lambda_small_r}(a), corresponds to Model 1 as shown in Fig.~\ref{fig_model1_OM_gap}(e).
Introducing infinitesimal pumping and loss induces a transition to Type 1 spectrum with a finite Liouvillian gap, as shown in Fig.~\ref{fig_model2_lambda_small_r}(b), where $\Delta_\mathrm{OM}>\Delta_L>0$.
For $r \geq 0.1$, both gaps equalize and remain positive, $\Delta_\mathrm{OM}=\Delta_L>0$, indicative of Type 2 spectrum as seen in Figs.~\ref{fig_model2_lambda_small_r}(c) and (d).
Thus, this analysis reveals the existence of two spectral types, Type 1 and Type 2, within the normal fluid phase of Model 2.

\section{Exact spectra}
\label{sec_exact_spectra}

In the previous section, we examined the Liouvillian spectrum using the mean-field approximation.
This section focuses on the exact Liouvillian spectrum of Model 1, derived through numerical diagonalization of the original Liouvillian.
It is important to note that true long-range order is absent in one-dimensional systems and typically manifests in three dimensions or higher.
Thus, the spectral structure corresponds to that depicted in Fig.~\ref{fig_model1_OM_gap}(e) across all parameter values.
Nevertheless, in finite one-dimensional systems, significant changes in spectral structure are anticipated with varying $\gamma$, highlighting indications of a potential phase transition.
Specifically, consider the correlation length $\xi$ of the one-dimensional system, which is the characteristic length scale of phase coherence.
Note that for Model 1, $\xi$ diverges as both the interaction strength $U$ and the dephasing rate $\gamma$ approach zero because a BEC state is realized at $U=0$ and $\gamma=0$ [see Fig.~\ref{fig_model1_OM_gap}(a)].
If the system size $L$ is smaller than or comparable to $\xi$, the mean-field results presented in Sec.~\ref{sec_mean_field_spectra} may approximately apply.
In such cases, the spectra of the system can exhibit behaviors similar to those predicted by the mean-field approximation.

\begin{figure}
\centering
\includegraphics[width=0.45\textwidth]{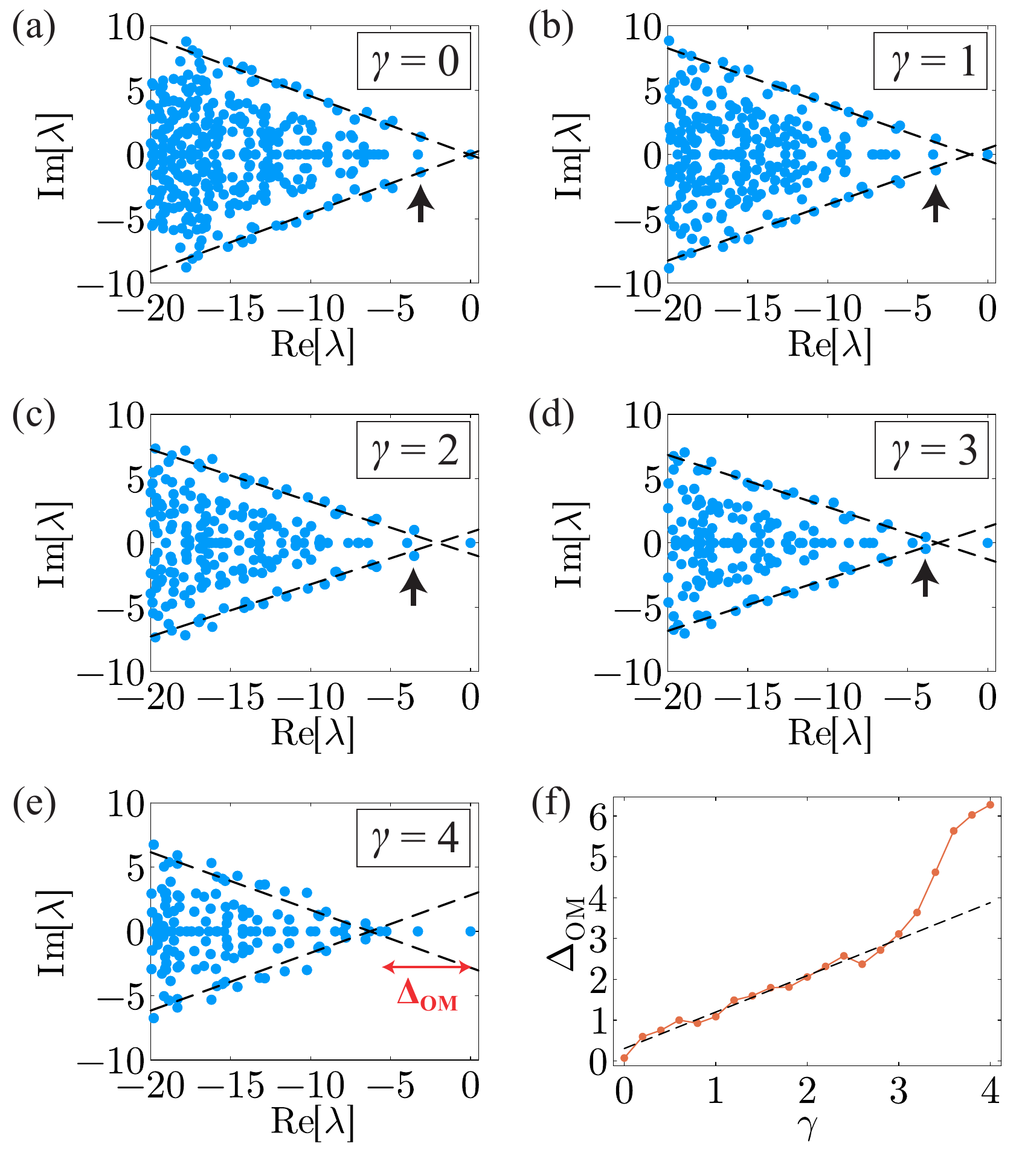}
\caption{Exact Liouvillian spectra for Model 1 with a system size $L=6$ and a particle number $N=3$.
The dephasing rates are (a) $\gamma=0$, (b) $\gamma=1$, (c) $\gamma=2$, (d) $\gamma=3$, and (e) $\gamma=4$.
Parameters are set to $\kappa=2$ and $U=2$.
The dashed lines represent the spectral edges, intersecting at points indicative of the OM gap $\Delta_\mathrm{OM}$.
The arrows indicate a pair of complex eigenvalues near the origin, which collide around $\gamma=3$.
Panel (f) displays $\Delta_\mathrm{OM}$ versus $\gamma$, with the dashed line representing a least-squares fit for the range $[0,2]$.
Note that $\Delta_\mathrm{OM}$ is derived from the intersection point of the edge of spectra, rather than using the original definition provided in Eq.~\eqref{def_Delta_OM}.
A rapid increase in $\Delta_\mathrm{OM}$ around $\gamma=3$ signals a potential phase transition.}
\label{fig_model1_spec_exact}
\end{figure}

Figure \ref{fig_model1_spec_exact} displays the exact spectra for Model 1 with a system size $L=6$ and a particle number $N=3$.
In the thermodynamic limit, the density of eigenvalues is expected to increase, leading to a continuum spectrum with two-dimensional support, contrasting with the one-dimensional structure of the mean-field spectrum.
Nevertheless, the mean-field spectrum is anticipated to reflect the edge structure of the exact spectrum.
In Figs.~\ref{fig_model1_spec_exact}(a)-(e), the behavior of the spectral edge, marked by dashed lines, resembles that observed in the mean-field spectrum shown in Fig.~\ref{fig_model1_OM_gap}(e) (see Appendix \ref{appendix_edge_detection} for edge detection methodology).
The intersection of these edge lines provides an approximate value of the OM gap $\Delta_\mathrm{OM}$.
Figure \ref{fig_model1_spec_exact}(f) plots $\Delta_\mathrm{OM}$ as a function of $\gamma$, showing a linear increase at low $\gamma$ values and a rapid ascent around $\gamma=3$, corresponding to the collision of a pair of complex eigenvalues near the origin, as indicated by arrows in Figs.~\ref{fig_model1_spec_exact}(a)-(d).
This rapid increase in $\Delta_\mathrm{OM}$ may indicate a potential phase transition in higher dimensions.
Accurate analysis of such singular behavior in the OM gap would require exact diagonalization of three-dimensional systems, which is currently beyond the scope of existing numerical techniques.

The real-complex transition in the eigenvalues of the most slowly-decaying mode, shown in Fig.~\ref{fig_model1_spec_exact}, is similarly observed in the XX chain with dephasing \cite{Znidaric-15}.
Specifically, a critical dephasing rate $\gamma_c$ exists, below which the eigenvalues of the most slowly-decaying mode are complex, and above which they become real.
At the critical dephasing rate $\gamma_c$, the most slowly-decaying mode transitions from a coherent mode, characterized by a density matrix with significant off-diagonal elements, to an incoherent mode, where the density matrix becomes nearly diagonal \cite{Haga-23}.
Importantly, the steady state of the XX chain with dephasing corresponds to the infinite-temperature state, implying the absence of phase transitions, and $\gamma_c$ is inversely proportional to the system size $L$, approaching zero in the limit of infinite system size.

It is important to address whether the closing of the OM gap is always accompanied by real-complex transitions of eigenvalues.
Let us consider the nonzero eigenvalue $\lambda^*$ that has the largest real part, such that $\Delta_L=|\mathrm{Re}[\lambda^*]|$.
There are two possible scenarios for the closing of the OM gap:
\begin{enumerate}
\item In the OM gapped phase, $\lambda^*$ is real, and at the phase transition of the steady state, it acquires a nonzero complex part through a collision with another real eigenvalue.
In this scenario, the phase transition point coincides with an exceptional point of the Liouvillian, where a real-complex transition of $\lambda^*$ occurs.
\item The closing of the OM gap occurs due to the shift of complex eigenvalues toward the positive direction of the real axis.
Here, $\lambda^*$, the eigenvalues with the largest real part, changes from a real one to another complex eigenvalue without a real-complex transition.
\end{enumerate}
At this stage, it is challenging to definitively determine which scenario applies to our case. 
The exact diagonalization of the one-dimensional system indicates a real-complex transition of $\lambda^*$, as shown in Fig.~\ref{fig_model1_spec_exact}, supporting the first scenario.
Conversely, the spectra obtained from the mean-field approximation seems to align with the second scenario (not shown in the figures), suggesting no real-complex transition during the closing of the OM gap.
However, it is important to note that the macroscopic dynamics of the system near the transition point remain unaffected by which scenario is valid.

\section{Relaxation dynamics}
\label{sec_relaxation_dynamics}

\begin{figure}
\centering
\includegraphics[width=0.45\textwidth]{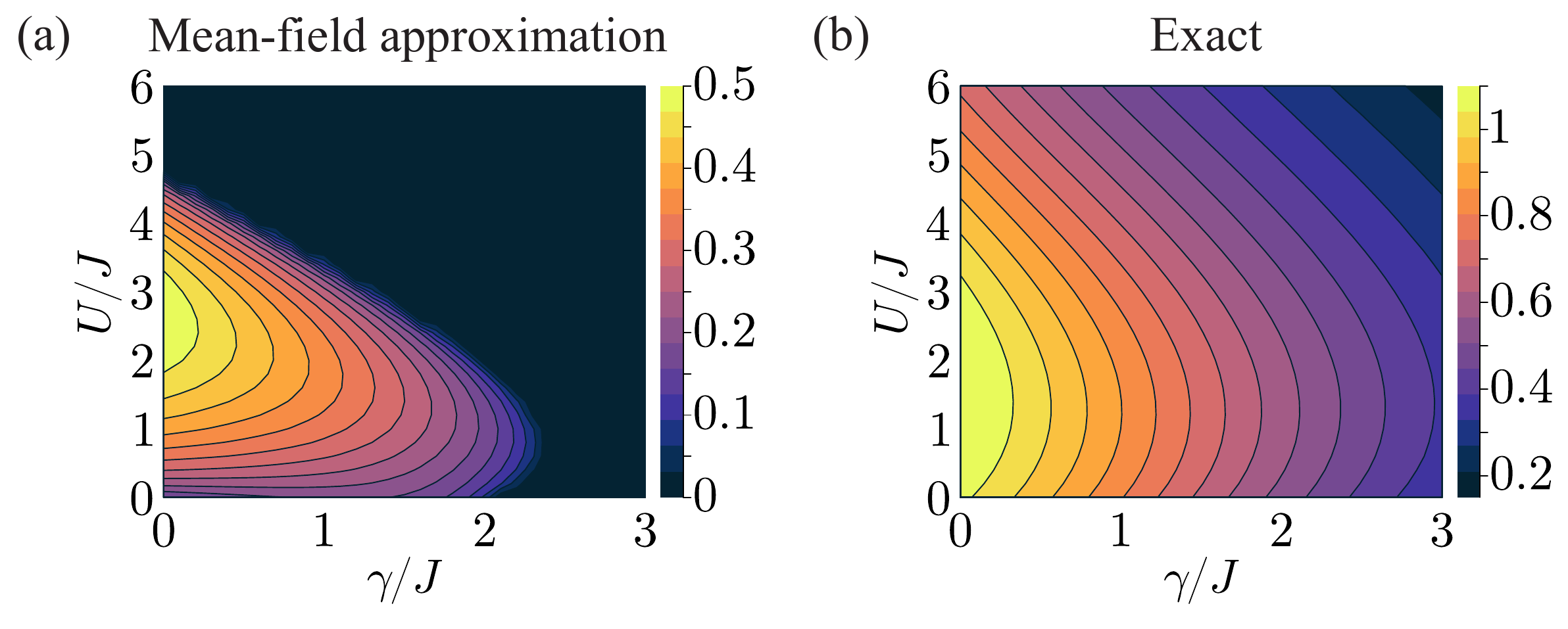}
\caption{Frequency of the most slowly-decaying mode in Model 1.
(a) Frequency of the density relaxation as a function of $\gamma/J$ and $U/J$ under the mean-field approximation.
Parameters include a particle density of $N/L=0.5$, a system size of $L=16$, and $\kappa=1$.
In the superfluid phase, the frequency of the most slowly-decaying mode is nonzero.
(b) Frequency of the density relaxation as a function of $\gamma/J$ and $U/J$, obtained from direct integration of the quantum master equation.
The calculations are performed for a system size of $L=6$ and a particle number of $N=3$.}
\label{fig_model1_relaxation}
\end{figure}

In this section, we discuss the relationship between the OM gap and the dynamical behavior of the system. 
Typically, the relaxation timescale towards the steady state is determined by the inverse of the Liouvillian gap $\Delta_L$. 
Therefore, a closure of $\Delta_L$ indicates a divergence in the relaxation timescale, a phenomenon observed in the phase transition of Model 2. 
In contrast, for Model 1, where the Liouvillian gap remains closed in both phases, no singular behavior in the relaxation timescale is anticipated at the transition point.
This prompts a question how the closure of the OM gap influences the relaxation dynamics of the system.

 In this analysis, we focus on the nonzero eigenvalue $\lambda^*$ that possesses the largest real part, such that $\Delta_L = |\mathrm{Re}[\lambda^*]|$.
 As depicted in Fig.~\ref{fig_model1_OM_gap}(c), when the OM gap is closed, $\mathrm{Im}[\lambda^*] \neq 0$, indicating that the most slowly-decaying mode exhibits oscillations.
In contrast, when the OM gap is open, as shown in Fig.~\ref{fig_model1_OM_gap}(e), $\mathrm{Im}[\lambda^*] = 0$ and the most slowly-decaying mode exhibits purely exponential decay.
Thus, the phase transition in Model 1 is characterized by a transition from oscillatory to exponential decay of the most slowly-decaying mode.
 
We present numerical results of relaxation dynamics in Model 1.
Considering the diffusive nature of density relaxation, the most slowly-decaying mode is characterized by large-scale density modulation, with a wavelength comparable to system size $L$.
Thus, we focus on the relaxation dynamics of an initial state with large-scale density modulation.
In the mean-field simulations, the initial state is prepared as follows.
The initial density matrix $\tilde{\rho}_j$ at site $j$ is given by $\tilde{\rho}_j = \ket{\psi_j}\bra{\psi_j}$, where $\ket{\psi_j}$ is the coherent state defined by Eq.~\eqref{coherent_state}. 
The initial order parameter $\psi_j$ is given by
\begin{equation}
\psi_j^2 = \bar{n} \left[ 1 + \delta \cos \left( \frac{2\pi j}{L} \right) \right],
\end{equation}
where $\delta$ denotes the amplitude of the initial density modulation, and $\bar{n}$ is the mean particle density.
The particle density at site $j$ and time $t$, $n_j(t)=\mathrm{tr}[b_j^\dag b_j \tilde{\rho}_j(t)]$, is used to define the amplitude of particle density modulation:
\begin{equation}
\delta n(t) = \sum_{j=1}^L n_j(t) \cos \left( \frac{2\pi j}{L} \right).
\label{density_modulation}
\end{equation}
The time evolution of $\delta n(t)$ is fitted to a function of $Ae^{-\Gamma t} \cos(\Omega t)$, where $\Gamma$ and $\Omega$ approximate the decay rate and frequency of the most slowly-decaying mode, respectively, $|\mathrm{Re}[\lambda^*]| \simeq \Gamma$, $|\mathrm{Im}[\lambda^*]| \simeq \Omega$.
In Fig.~\ref{fig_model1_relaxation}(a), the frequency $\Omega$ of the most slowly-decaying mode is displayed as a function of $\gamma/J$ and $U/J$.
In the bright region, the density perturbation exhibits oscillations, whereas in the dark region, it decays exponentially.
By comparing with Fig.~\ref{fig_model1_OM_gap}(a), it is observed that the phase boundary between the superfluid and normal fluid coincides with the transition from oscillatory to exponential relaxations.

For a small system, we can calculate the exact relaxation dynamics by directly solving the original quantum master equation \eqref{master_eq_general}.
For a system size of $L=6$ and a particle number of $N=3$, the initial state is set as $\ket{1, 1, 1, 0, 0, 0}$, dividing the system equally between filled and empty sites.
The time evolution of the many-body density matrix $\rho(t)$ is calculated, and the amplitude of the density modulation $\delta n(t)$, as defined by Eq.~\eqref{density_modulation}, is determined.
The frequency of the most slowly-decaying mode derived from $\delta n(t)$ is displayed in Fig.~\ref{fig_model1_relaxation}(b), showing a qualitatively similar trend to the mean-field calculations presented in Fig.~\ref{fig_model1_relaxation}(a).
However, it is important to note that true long-range order does not exist in one dimension, and thus the exact time evolution results displayed in Fig.~\ref{fig_model1_relaxation}(b) do not converge to the mean-field results in panel (a) as the system size increases.

\section{Conclusions and discussion}
\label{sec_conclusions}

In this paper, we introduce a novel spectral gap, termed the oscillating-mode (OM) gap, to characterize phase transitions in open quantum many-body systems. 
Traditionally, the Liouvillian gap, defined as the smallest decay rate among all non-steady eigenmodes, has been extensively used to indicate dissipative quantum phase transitions (DQPTs). 
The OM gap, however, is defined as the smallest decay rate among oscillating eigenmodes with nonzero imaginary parts of their eigenvalues. 
This distinction allows for a more nuanced classification of open quantum phases into four distinct spectral classes, as illustrated in Fig.~\ref{fig_spectrum_class}. 
Our analysis within dissipative Bose systems, referred to as Model 1 and Model 2, demonstrates the necessity of both the Liouvillian and OM gaps for accurately describing the phases and phase transitions within these systems.

Model 1 exhibits the strong $U(1)$ symmetry, ensuring conservation of particle number. 
This model demonstrates a symmetry-broken phase at low dephasing rates and interaction strengths. 
The Liouvillian gap $\Delta_L$ closes in both ordered and disordered phases, which limits its utility in distinguishing these phases. 
In contrast, transitions between phases can be effectively characterized by the closing or opening of the OM gap $\Delta_\mathrm{OM}$. 
Additionally, the closure of $\Delta_\mathrm{OM}$ in the ordered phase is marked by a dynamical transition from exponential to oscillatory relaxations of inhomogeneous states with large-scale density modulation.

In Model 2, the particle number is not conserved due to the presence of pumping and loss mechanisms.
This model also presents a symmetry-broken phase under conditions of high pumping rates and low interaction strengths. 
The Liouvillian gap $\Delta_L$ closes in the ordered phase and opens in the disordered phase, accordance with the traditional picture of quantum phase transitions.
By contrast, the OM gap $\Delta_\mathrm{OM}$ closes solely at the phase transition point. 
Notably, the disordered phase in Model 2 bifurcates into two distinct phases characterized either by $\Delta_\mathrm{OM} > \Delta_L$ or $\Delta_\mathrm{OM} = \Delta_L$.

Our analysis suggests that not only the Liouvillian gap $\Delta_L$, but also the OM gap $\Delta_{\mathrm{OM}}$ closes when the steady state undergoes spontaneous symmetry breaking.
For example, in Model 2, $\Delta_\mathrm{OM}$ closes exclusively at the transition point, as shown in Fig.~\ref{fig_model2_OM_gap}(d).
This raises an intriguing question: Can spontaneous symmetry breaking occur {\it without} the closure of the OM gap?
Note that the spectral transition from Type 3 to Type 1 in Fig.~\ref{fig_model2_lambda_small_r} is not associated with spontaneous symmetry breaking in the steady state.
If such a DQPT exists, it would represent a form of transition that occurs without exceptional points near the steady state, since all eigenvalues close to the origin remain real across the transition point.

We mention the relationship between spectral classes and system's symmetry.
In open quantum system described by a Liouvillian with the strong $U(1)$ symmetry, total particle number or spin component is conserved through time evolution.
It is conjectured that for translation-invariant Liouvillians incorporating only local interactions and possessing local conserved quantities, the Liouvillian gap $\Delta_L$ vanishes in the thermodynamic limit.
This phenomenon is explained by the limited propagation speed of local perturbations, constrained by the Lieb-Robinson velocity \cite{Lieb-72, Poulin-10}.
Consequently, an initial state characterized by localized particle density requires a relaxation time that is, at least, proportional to the system size to evolve towards a uniform steady state.
Given that the relaxation time is inversely proportional to $\Delta_L$ \cite{footnotes}, it follows that $\Delta_L \to 0$ in the thermodynamic limit.
Therefore, a translation-invariant system with local conserved quantities will typically exhibit either Type 3 or Type 4 spectra shown in Fig.~\ref{fig_spectrum_class}.
Moreover, phase transitions in such systems should be marked by the closing of the OM gap, indicating that DQPTs associated with the closure of the OM gap are a prevalent feature in open quantum many-body physics.

From the viewpoint of the universality of critical phenomena, it is natural to introduce the critical exponent associated with the OM gap.
In classical critical phenomena theory, as one approaches the critical point, the correlation length $\xi$ diverges as $\xi \propto |t|^{-\nu}$, where $t$ is a dimensionless measure of the distance from the critical point.
The typical timescale $\tau$ for the decay of fluctuations also diverges as 
$\tau \propto \xi^z \propto |t|^{- \nu z}$, where $z$ is the dynamical critical exponent.
Given that $\tau$ is identical to the inverse of the Liouvillian gap $\Delta_L$, it follows that $\Delta_L$ vanishes as
\begin{equation}
\Delta_L \propto \xi^{-z} \propto |t|^{\nu z},
\end{equation}
near the critical point.
Similarly, for the OM gap, it is assumed that
\begin{equation}
\Delta_\mathrm{OM} \propto \xi^{-\tilde{z}} \propto |t|^{\nu \tilde{z}},
\end{equation}
with a distinct exponent $\tilde{z}$.
In Model 1 and Model 2 studied in this work, we numerically confirm that $\nu z = \nu \tilde{z} = 1$ within the mean-field approximation.
Nonetheless, it is plausible that quantum fluctuations not captured by this approximation could yield nontrivial exponents.
The introduction of a new scaling law for the OM gap may reveal a novel universality class in DQPTs, distinct from those observed in closed quantum many-body systems.

We here discuss the relationship between our findings and the concept of continuous time crystals (CTCs) in open quantum systems \cite{Iemini-18, Lledo-19, Piccitto-21, Prazeres-21, Kongkhambut-22, Hajdusek-22, Krishna-23}.
A CTC arises from the breaking of continuous time-translation symmetry and is characterized by an oscillating steady state in the thermodynamic limit.
Importantly, the emergence of a CTC implies the vanishing of the OM gap. 
This vanishing indicates that the relaxation time of some oscillating modes diverges, allowing persistent oscillations in the thermodynamic limit.
However, it is crucial to note that the vanishing of the OM gap does not necessarily imply the emergence of a CTC.
Let $\tilde{\lambda}$ be the eigenvalue with the largest real part among those with nonzero imaginary parts.
The OM gap is defined as $\Delta_\mathrm{OM} = |\mathrm{Re}[\tilde{\lambda}]|$.
Additionally, we can define a gap along the imaginary axis, $\Delta_\mathrm{OM}' = |\mathrm{Im}[\tilde{\lambda}]|$.
In the thermodynamic limit, an OM gapless phase, which is characterized by Type 4 spectrum in Fig.~\ref{fig_spectrum_class}, can be classified into two cases:
\begin{enumerate}
\item $\Delta_\mathrm{OM} = 0$ and $\Delta_\mathrm{OM}' = 0$.
\item $\Delta_\mathrm{OM} = 0$ and $\Delta_\mathrm{OM}' > 0$.
\end{enumerate}
The first case, where both the real and imaginary parts of $\tilde{\lambda}$ vanish, corresponds to the scenario in our model, indicating that no CTC phase emerges. 
The second case, where only the real part vanishes while the imaginary part remains nonzero, corresponds to the CTC phase.
Thus, the concept of OM gap might be useful for characterizing the CTC phase.

Finally, we note that the concept of the OM gap is also relevant to classical stochastic systems governed by the standard master equation. 
Generally, the transition rate matrix of the master equation is non-Hermitian, leading to a complex spectrum. 
Thus, the spectral classification illustrated in Fig.~\ref{fig_spectrum_class} is applicable to these cases as well. 
There exists a broad range of nonequilibrium stochastic models, including driven lattice gases \cite{Derrida-98}, reaction-diffusion processes \cite{Hinrichsen-00}, and active matters \cite{Marchetti-13}, that display phase transitions in their steady states.
A critical avenue for future research is to elucidate the spectral structures that underpin these well-documented nonequilibrium phase transitions.

\begin{acknowledgments}
This work was supported by JSPS KAKENHI Grant Number JP22K13983.
\end{acknowledgments}

\appendix

\section{General properties of Liouvillian eigenmodes}
\label{appendix_general_properties_of_Liouvillian_eigenmodes}

This summary outlines the general properties of the eigenmodes and eigenvalues of a Liouvillian, defined in Eq.~\eqref{eigen_eq}.
\begin{enumerate}
\item {\it Left Eigenmodes:} The left eigenmodes $\{ \rho_{\alpha}' \}$ satisfy
\begin{equation}
\mathcal{L}^{\dag}(\rho_{\alpha}')=\lambda_{\alpha}^* \rho_{\alpha}' \quad (\alpha=0, 1, ... , D^2-1),
\end{equation}
where $\mathcal{L}^{\dag}$ operates as:
\begin{equation}
\mathcal{L}^{\dag}(A) = -i[A, H] + \sum_{\nu} \left( L_{\nu}^{\dag} A L_{\nu} - \frac{1}{2} \{ L_{\nu}^{\dag}L_{\nu}, A \} \right).
\end{equation}
Given that $\mathcal{L}$ is non-Hermitian, the right eigenmode $\rho_{\alpha}$ and the left eigenmode $\rho_{\alpha}'$ differ.
The right and left eigenmodes corresponding to different eigenvalues are orthogonal: $\mathrm{tr}[\rho_{\alpha}'^{\dag} \rho_{\beta}] = 0 \: (\lambda_\alpha \neq \lambda_\beta)$.

\item {\it Diagonalizability:} The Liouvillian is diagonalizable except at points in the parameter space, i.e., exceptional points.
In such cases, the eigenmodes form a basis of the operator space, allowing any operator $O$ to be uniquely expressed as:
\begin{equation}
O = \sum_{\alpha=0}^{D^2-1} c_\alpha \rho_{\alpha}, \quad c_\alpha = \frac{\mathrm{tr}[\rho_{\alpha}'^{\dag} O]}{\mathrm{tr}[\rho_{\alpha}'^{\dag} \rho_{\alpha}]},
\end{equation}
where the expression of $c_\alpha$ is derived from the orthogonality of the right and left eigenmodes.

\item {\it Real Parts and Zero Modes:} Zero modes correspond to steady states.
For other modes, the real parts of $\lambda_{\alpha}$ are negative, ensuring relaxation toward steady states:
\begin{equation}
\lim_{t \to \infty} \rho(t) = \lim_{t \to \infty} e^{\mathcal{L}t} \rho(0) = \rho_{\mathrm{ss}}.
\end{equation}

\item {\it Liouvillian Gap:} The Liouvillian gap is defined by $\Delta_L=|\mathrm{Re}[\lambda_1]|$, provided the steady state is unique and the eigenvalues are sorted accordingly, $0 = |\mathrm{Re}[\lambda_0]| < |\mathrm{Re}[\lambda_1]| \leq \cdots \leq |\mathrm{Re}[\lambda_{D^2-1}]|$.
$\Delta_L$ is also referred to as the asymptotic decay rate \cite{Kessler-12}.

\item {\it Trace of Eigenmodes:} The trace of any eigenmode associated with a nonzero eigenvalue is zero, reflecting the trace-preserving property of the Liouvillian: $\mathrm{tr}[\mathcal{L}(\rho)] = 0$.
Thus, eigenmodes with nonzero eigenvalues are not physical states.

\item {\it Hermitian Properties:} If $\mathcal{L}(\rho_{\alpha}) = \lambda_{\alpha} \rho_{\alpha}$, then $\mathcal{L}(\rho_{\alpha}^{\dag}) = \lambda_{\alpha}^* \rho_{\alpha}^{\dag}$, which follows from $[\mathcal{L}(\rho_{\alpha})]^{\dag} = \mathcal{L}(\rho_{\alpha}^{\dag})$.
This implies that the Liouvillian spectrum on the complex plane is symmetric with respect to the real axis.
If $\rho_{\alpha}$ is Hermitian, $\lambda_{\alpha}$ is real.
\end{enumerate}

\section{Microscopic Hamiltonian}
\label{appendix_microscopic_Hamiltonian}

In this Appendix, we discuss the microscopic Hamiltonians leading to the dissipation mechanisms given by Eqs.~\eqref{L_bond}-\eqref{L_2P_loss}.
Firstly, the bond dissipation given by Eq.~\eqref{L_bond} stabilizes a superfluid state with phase coherence and can be realized in ultracold atoms in a two-band optical lattice immersed in a large BEC \cite{Diehl-08}.
Spatially-modulated Raman laser drives the specific superposition of atoms on neighboring sites into the upper band, which decays to the lower band by emitting a Bogoliubov quasi-particle. 
This process is described by a two-band Bose-Hubbard model with a laser interaction term, a reservoir of Bogoliubov modes, and coupling terms between the atoms and the Bogoliubov modes.
The explicit form of the microscopic Hamiltonian is given in Ref.~\cite{Diehl-08}.

The dephasing given by Eq.~\eqref{L_dephasing} describes the loss of phase coherence without energy dissipation, arising from a system under a fluctuating external field.
The microscopic Hamiltonian is given by
\begin{equation}
\mathcal{H}(t) = H + \sum_\nu \xi_\nu(t) L_\nu,
\end{equation}
where $\xi_\nu(t)$ denotes Gaussian white noise with $\overline{\xi_\nu(t)}=0$ and $\overline{\xi_\mu(t)\xi_\nu(t')} = \delta_{\mu \nu} \delta(t-t')$.
The Hermitian operators $L_\nu$ represent the interactions with the external field.
The master equation for the density matrix $\rho(t)$ is then given by Eq.~\eqref{master_eq_general}.
This derivation can be found in Refs.~\cite{Pichler-13} and \cite{Stannigel-14}.
Coupling the density operator $b_j^\dag b_j$ to independent fluctuating fields $\xi_j(t)$ leads to the local jump operator in Eq.~\eqref{L_dephasing}.
Note that this formalism cannot yield a master equation with non-Hermitian jump operators $L_\nu$.

The particle gain and loss processes described by Eqs.~\eqref{L_pumping} and \eqref{L_1P_loss} are modeled by the following Hamiltonian:
\begin{equation}
H_\mathrm{tot} = H + H_E + H_\mathrm{int},
\end{equation}
where $H$ is the Hamiltonian \eqref{Hamiltonian} of the system, $H_E$ is that of an environment, and $H_\mathrm{int}$ is a coupling between them.
The environment consists of independent baths coupled to every site $j$:
\begin{equation}
H_E = \sum_j \sum_k \epsilon_k a_{j,k}^\dag a_{j,k},
\end{equation}
where $a_{j,k}^\dag$ and $a_{j,k}$ are creation and annihilation operators for particles with orbital degrees of freedom $k$ in a bath coupled to site $j$.
The coupling between the system and environment is described by
\begin{equation}
H_\mathrm{int} = \sum_j \sum_k t_k b_j^\dag a_{j, k} + \mathrm{H.c.},
\end{equation}
where $t_k$ is the tunneling amplitude for a particle entering or leaving site $j$.
The derivation of the master equation is detailed in Ref.~\cite{Kleinherbers-20}.
It is important to note that the resulting jump operators $L_j$ are nonlocal due to particle hopping.
In particular, this nonlocal effect becomes prominent in the zero-temperature limit.
However, at sufficiently high temperatures, it has been shown that the jump operators $L_j$ can be approximated by the local ones \cite{Kleinherbers-20}, as in Eqs.~\eqref{L_pumping} and \eqref{L_1P_loss}.
The two particle loss term given by Eq.~\eqref{L_2P_loss} can be derived by modifying $H_\mathrm{int}$ to include two-particle tunneling.

\section{Mean-field master equation}
\label{appendix_mean_field_master_equation}

In this Appendix, we present the explicit form of the mean-field equation \eqref{master_eq_mf} and its linearized version in Eq.~\eqref{master_eq_mf_linearized}.
These formulations are partially based on the work in Ref.~\cite{Tomadin-11}.

\begin{widetext}
The right-hand side of Eq.~\eqref{master_eq_mf} is written as
\begin{equation}
\mathcal{L}_\mathrm{MF}( \tilde{\rho}_j; \tilde{\rho}_{j-1}, \tilde{\rho}_{j+1}) = \mathcal{L}_\mathrm{MF}^H ( \tilde{\rho}_j; \tilde{\rho}_{j-1}, \tilde{\rho}_{j+1}) + \mathcal{L}_\mathrm{MF}^b ( \tilde{\rho}_j; \tilde{\rho}_{j-1}, \tilde{\rho}_{j+1}) + \sum_{\nu = d, p, l, t} \mathcal{L}_\mathrm{MF}^\nu ( \tilde{\rho}_j),
\end{equation}
where each term represents the contributions from unitary evolution, bond dissipation, dephasing, pumping, one-particle loss, and two-particle loss, respectively.
The unitary contribution is expressed as
\begin{equation}
\mathcal{L}_\mathrm{MF}^H ( \tilde{\rho}_j; \tilde{\rho}_{j-1}, \tilde{\rho}_{j+1}) = - i [h_j, \tilde{\rho}_j],
\end{equation}
\begin{equation}
h_j = - J \left( \langle b_{j+1} \rangle b_j^\dag + \langle b_{j+1}^\dag \rangle b_j + \langle b_{j-1} \rangle b_j^\dag + \langle b_{j-1}^\dag \rangle b_j \right) + \frac{U}{2} b_j^\dag b_j^\dag b_j b_j - \mu b_j^\dag b_j,
\end{equation}
where $\langle b_{j\pm1} \rangle = \mathrm{tr}[b_{j\pm1} \tilde{\rho}_{j\pm1}]$.
The bond dissipation term is given by
\begin{equation}
\mathcal{L}_\mathrm{MF}^b ( \tilde{\rho}_j; \tilde{\rho}_{j-1}, \tilde{\rho}_{j+1}) = \kappa \sum_{r, s = 1}^4 \left[ A_j^r \ \tilde{\rho}_j \ A_j^{s \dag} - \frac{1}{2} \left\{A_j^{s \dag} A_j^r , \tilde{\rho}_j \right\} \right] \left( \Gamma_{j+1}^{rs} + \Gamma_{j-1}^{rs} \right),
\end{equation}
where $A_j = (1, b_j^\dag, b_j, n_j)$ and a 4 by 4 matrix $\Gamma_j$ is given by
\begin{equation}
\Gamma_j = 
\begin{pmatrix}
	\langle n_j^2 \rangle & \langle b_j^\dag n_j \rangle & -\langle b_j n_j \rangle & - \langle n_j \rangle \\
	\langle n_j b_j \rangle  & \langle n_j \rangle & -\langle b_j^2 \rangle & -\langle b_j\rangle \\
	- \langle n_j b_j^\dag \rangle & -\langle b_j^{\dag 2} \rangle & \langle n_j \rangle +1 & \langle b_j^\dag \rangle \\
	-\langle n_j \rangle & -\langle b_j^\dag\rangle & \langle b_j \rangle & 1 \\
\end{pmatrix}.
\end{equation}
The contributions from dephasing, pumping, one-particle loss, and two-particle loss are represented as
\begin{equation}
\mathcal{L}_\mathrm{MF}^d ( \tilde{\rho}_j) = \gamma \left( n_j \tilde{\rho}_j n_j - \frac{1}{2} \left\{ n_j^2, \tilde{\rho}_j \right\} \right),
\label{appendix_L_MF_d}
\end{equation}
\begin{equation}
\mathcal{L}_\mathrm{MF}^p ( \tilde{\rho}_j) = r_p \left( b_j^\dag \tilde{\rho}_j b_j - \frac{1}{2} \left\{ b_j b_j^\dag , \tilde{\rho}_j \right\} \right),
\label{appendix_L_MF_p}
\end{equation}
\begin{equation}
\mathcal{L}_\mathrm{MF}^l ( \tilde{\rho}_j) = r_l \left( b_j \tilde{\rho}_j b_j^\dag - \frac{1}{2} \left\{ b_j^\dag b_j , \tilde{\rho}_j \right\} \right),
\label{appendix_L_MF_l}
\end{equation}
\begin{equation}
\mathcal{L}_\mathrm{MF}^t ( \tilde{\rho}_j) = r_t \left( b_j b_j \tilde{\rho}_j b_j^\dag b_j^\dag - \frac{1}{2} \left\{ b_j^\dag b_j^\dag b_j b_j , \tilde{\rho}_j \right\} \right).
\label{appendix_L_MF_t}
\end{equation}

For a sufficiently long time integration of the mean-field master equation, a steady state $\tilde{\rho}_\mathrm{ss}$ is achieved.
The linearized master equation for a small perturbation $\delta \tilde{\rho}_j = \tilde{\rho}_j - \tilde{\rho}_\mathrm{ss}$ is given by
\begin{align}
\partial_t \delta \tilde{\rho}_j =& - i [h_{\mathrm{ss}, j}, \delta \tilde{\rho}_j] - i [\delta h_j, \tilde{\rho}_\mathrm{ss}] + F_b^1 ( \delta \tilde{\rho}_j) + F_b^2 ( \delta \tilde{\rho}_{j-1}, \delta \tilde{\rho}_{j+1}) + \sum_{\nu = d, p, l, t} \mathcal{L}_\mathrm{MF}^\nu ( \delta \tilde{\rho}_j),
\end{align}
Here, $h_{\mathrm{ss}, j}$ denotes the local Hamiltonian for the steady state,
\begin{equation}
h_{\mathrm{ss}, j} = - J \left( \langle b_{j+1} \rangle_\mathrm{ss} b_j^\dag + \langle b_{j+1}^\dag \rangle_\mathrm{ss} b_j + \langle b_{j-1} \rangle_\mathrm{ss} b_j^\dag + \langle b_{j-1}^\dag \rangle_\mathrm{ss} b_j \right) + \frac{U}{2} b_j^\dag b_j^\dag b_j b_j - \mu b_j^\dag b_j,
\end{equation}
and $\delta h_j$ represents the perturbed local Hamiltonian,
\begin{equation}
\delta h_j = - J \left( \langle \delta b_{j+1} \rangle b_j^\dag + \langle \delta b_{j+1}^\dag \rangle b_j + \langle \delta b_{j-1} \rangle b_j^\dag + \langle \delta b_{j-1}^\dag \rangle b_j \right),
\end{equation}
where $\langle O_j \rangle_\mathrm{ss} = \mathrm{tr}[ \tilde{\rho}_\mathrm{ss} O_j]$ and $\langle \delta O_j \rangle = \mathrm{tr}[ \delta \tilde{\rho}_j O_j]$.
The contributions from the bond dissipation are given by
\begin{equation}
F_b^1 ( \delta \tilde{\rho}_j) = \kappa \sum_{r, s = 1}^4 \left[ A_j^r \ \delta \tilde{\rho}_j \ A_j^{s \dag} - \frac{1}{2} \left\{ A_j^{s \dag} A_j^r , \delta \tilde{\rho}_j \right\} \right] \left( \Gamma_{\mathrm{ss}, j+1}^{rs} + \Gamma_{\mathrm{ss}, j-1}^{rs} \right),
\end{equation}
\begin{equation}
F_b^2 ( \delta \tilde{\rho}_{j-1}, \delta \tilde{\rho}_{j+1}) = \kappa \sum_{r, s = 1}^4 \left[ A_j^r \ \tilde{\rho}_\mathrm{ss} \ A_j^{s \dag} - \frac{1}{2} \left\{ A_j^{s \dag} A_j^r , \tilde{\rho}_\mathrm{ss} \right\} \right] \left( \delta \Gamma_{j+1}^{rs} + \delta \Gamma_{j-1}^{rs} \right),
\end{equation}
where matrices $\Gamma_{\mathrm{ss}, j}$ and $\delta \Gamma_j$ read
\begin{equation}
\Gamma_{\mathrm{ss}, j} = 
\begin{pmatrix}
		\langle n_j^2 \rangle_\mathrm{ss} & \langle b_j^\dag n_j \rangle_\mathrm{ss} & -\langle b_j n_j \rangle_\mathrm{ss} & - \langle n_j \rangle_\mathrm{ss} \\
		\langle n_j b_j \rangle_\mathrm{ss} & \langle n_j \rangle_\mathrm{ss} & -\langle b_j^2 \rangle_\mathrm{ss} & -\langle b_j\rangle_\mathrm{ss} \\
		- \langle n_j b_j^\dag \rangle_\mathrm{ss} & -\langle b_j^{\dag 2} \rangle_\mathrm{ss} & \langle n_j \rangle_\mathrm{ss} +1 & \langle b_j^\dag \rangle_\mathrm{ss} \\
		-\langle n_j \rangle_\mathrm{ss} & -\langle b_j^\dag\rangle_\mathrm{ss} & \langle b_j \rangle_\mathrm{ss} & 1 \\
\end{pmatrix},
\quad
\delta \Gamma_j = 
\begin{pmatrix}
		\langle \delta (n_j^2) \rangle & \langle \delta (b_j^\dag n_j) \rangle & - \langle \delta (b_j n_j) \rangle & - \langle \delta n_j \rangle \\
		\langle \delta (n_j b_j) \rangle  & \langle \delta n_j \rangle & -\langle \delta (b_j^2) \rangle & - \langle \delta b_j \rangle \\
		- \langle \delta (n_j b_j^\dag) \rangle & -\langle \delta (b_j^{\dag 2}) \rangle & \langle \delta n_j \rangle & \langle \delta  b_j^\dag \rangle \\
		-\langle \delta n_j \rangle & -\langle \delta b_j^\dag \rangle & \langle \delta b_j \rangle & 0 \\
\end{pmatrix}.
\end{equation}
The contributions from dephasing, pumping, one-particle loss, and two-particle loss are obtained by replacing $\tilde{\rho}_j$ with $\delta \tilde{\rho}_j$ in Eq.~\eqref{appendix_L_MF_d}, \eqref{appendix_L_MF_p}, \eqref{appendix_L_MF_l}, \eqref{appendix_L_MF_t}.
\end{widetext}

\section{Edge detection for exact spectrum}
\label{appendix_edge_detection}

In this Appendix, we outline the method employed to detect the edges of exact spectra depicted in Fig.~\ref{fig_model1_spec_exact}.
Consider a set of two-dimensional vectors $\{ (x_i, y_i) \}_{i=1,...,N}$ representing the Liouvillian spectrum in the complex plane.
For each data point, the local density is calculated by
\begin{equation}
D_i = \sum_{j=1}^N K(x_i - x_j, y_i - y_j),
\end{equation}
where $K(x, y)$ is a Gaussian kernel defined as
\begin{equation}
K(x, y) = \frac{1}{2\pi \sigma^2} \exp \left[ - \frac{x^2 + y^2}{2\sigma^2} \right].
\end{equation}
A threshold $D_{\mathrm{th}}$ is set, and data points with $D_i < D_{\mathrm{th}}$ are classified as edge points. 
These points are then fitted with a straight line using least-squares fitting to define the spectral edge. 
In Fig.~\ref{fig_model1_spec_exact}, the standard deviation of the kernel is set at $\sigma=1$, and the threshold for local density is set at $D_{\mathrm{th}}=1.5$.


\begin{thebibliography}{99}
\bibitem{Hastings-04-1} M. B. Hastings, Locality in Quantum and Markov Dynamics on Lattices and Networks, Phys. Rev. Lett. {\bf 93}, 140402 (2004).

\bibitem{Hastings-04-2} M. B. Hastings, Lieb-Schultz-Mattis in higher dimensions, Phys. Rev. B {\bf 69}, 104431 (2004).

\bibitem{Hastings-06} M. B. Hastings and T. Koma, Spectral Gap and Exponential Decay of Correlations, Commun. Math. Phys. {\bf 265}, 781 (2006).

\bibitem{Hastings-07} M. B. Hastings, Entropy and entanglement in quantum ground states, Phys. Rev. B {\bf 76}, 035114 (2007).

\bibitem{Sachdev} S. Sachdev, {\it Quantum Phase Transitions} (Cambridge University Press, Cambridge, England, 2001).

\bibitem{Bloch-08-1} I. Bloch, J. Dalibard, and W. Zwerger, Many-body physics with ultracold gases, Rev. Mod. Phys. {\bf 80}, 885 (2008).

\bibitem{Bloch-08-2} I. Bloch, Quantum coherence and entanglement with ultracold atoms in optical lattices, Nature {\bf 453}, 1016 (2008).

\bibitem{Syassen-08} N. Syassen, D. M. Bauer, M. Lettner, T. Volz, D. Dietze, J. J. Garcia-Ripoll, J. I. Cirac, G. Rempe, and S. D{\"u}rr, Strong dissipation inhibits losses and induces correlations in cold molecular gases, Science {\bf 320}, 1329 (2008).

\bibitem{Bloch-12} I. Bloch, J. Dalibard, and S. Nascimb\'ene, Quantum simulations with ultracold quantum gases, Nat. Phys. {\bf 8}, 267 (2012).

\bibitem{Ritsch-13} H. Ritsch, P. Domokos, F. Brennecke, and T. Esslinger, Cold atoms in cavity-generated dynamical optical potentials, Rev. Mod. Phys. {\bf 85}, 553 (2013).

\bibitem{Lanyon-09} B. P. Lanyon {\it et al.,} Universal Digital Quantum Simulation with Trapped Ion, Science {\bf 334}, 57 (2009).

\bibitem{Barreiro-11} J. T. Barreiro, M. M{\"u}ller, P. Schindler, D. Nigg, T. Monz, M. Chwalla, M. Hennrich, C. F. Roos, P. Zoller, and R. Blatt, An open-system quantum simulator with trapped ions, Nature {\bf 470}, 486 (2011).

\bibitem{Blatt-12} R. Blatt and C. F. Roos, Quantum simulations with trapped ions, Nat. Phys. {\bf 8}, 277 (2012).

\bibitem{Hartmann-06} M. J. Hartmann, F. G. S. L. Brand\~ao, and M. B. Plenio, Strongly interacting polaritons in coupled arrays of cavities, Nat. Phys. {\bf 2}, 849 (2006).

\bibitem{Baumann-10} K. Baumann, C. Guerlin, F. Brennecke, and T. Esslinger, Dicke quantum phase transition with a superfluid gas in an optical cavity, Nature {\bf 464}, 1301 (2010).

\bibitem{Eichler-14} C. Eichler, Y. Salathe, J. Mlynek, S. Schmidt, and A. Wallraff, Quantum-Limited Amplification and Entanglement in Coupled Nonlinear Resonators, Phys. Rev. Lett. {\bf 113}, 110502 (2014).

\bibitem{Rodriguez-16} S. R. K. Rodriguez, A. Amo, I. Sagnes, L. Le Gratiet, E. Galopin, A. Lema\^itre, and J. Bloch, Interaction-induced hopping phase in driven-dissipative coupled photonic microcavities, Nat. Commun. {\bf 7}, 11887 (2016). 

\bibitem{Lindblad-76} G. Lindblad, On the generators of quantum dynamical semigroups, Commun. Math. Phys. {\bf 48}, 119 (1976).

\bibitem{Gorini-76} V. Gorini, A. Kossakowski, and E. C. G. Sudarshan, Completely positive dynamical semigroups of N-level systems, J. Math. Phys. {\bf 17}, 821 (1976).

\bibitem{Breuer} H. P. Breuer and F. Petruccione, {\it The Theory of Open Quantum Systems} (Oxford University Press, Oxford, 2002).

\bibitem{Rivas} A. Rivas and S. F. Huelga, {\it Open Quantum Systems} (Springer, New York, 2012).

\bibitem{Kessler-12} E. M. Kessler, G. Giedke, A. Imamoglu, S. F. Yelin, M. D. Lukin,  and J. I. Cirac, Dissipative phase transition in a central spin system, Phys. Rev. A {\bf 86}, 012116 (2012).

\bibitem{Honing-12} M. H\"oning, M. Moos, and M. Fleischhauer, Critical exponents of steady-state phase transitions in fermionic lattice models, Phys. Rev. A {\bf 86}, 013606 (2012).

\bibitem{Horstmann-13} B. Horstmann, J. I. Cirac, and G. Giedke, Noise-driven dynamics and phase transitions in fermionic systems, Phys. Rev. A {\bf 87}, 012108 (2013).

\bibitem{Casteels-16} W. Casteels, F. Storme, A. Le Boit\'{e}, and C. Ciuti, Power laws in the dynamic hysteresis of quantum nonlinear photonic resonators, Phys. Rev. A {\bf 93}, 033824 (2016).

\bibitem{Casteels-17} W. Casteels, R. Fazio, and C. Ciuti, Critical dynamical properties of a first-order dissipative phase transition, Phys. Rev. A {\bf 95}, 012128 (2017).

\bibitem{Fitzpatrick-17} M. Fitzpatrick, N. M. Sundaresan, A. C. Y. Li, J. Koch, and A. A. Houck, Observation of a Dissipative Phase Transition in a One-Dimensional Circuit QED Lattice, Phys. Rev. X {\bf 7}, 011016 (2017).

\bibitem{Vicentini-18} F. Vicentini, F. Minganti, R. Rota, G. Orso, and C. Ciuti, Critical slowing down in driven-dissipative Bose-Hubbard lattices, Phys. Rev. A {\bf 97}, 013853 (2018).

\bibitem{Minganti-18} F. Minganti, A. Biella, N. Bartolo, and C. Ciuti, Spectral theory of Liouvillians for dissipative phase transitions, Phys. Rev. A {\bf 98}, 042118 (2018).

\bibitem{Imamoglu-18} T. Fink, A. Schade, S. H\"ofling, C. Schneider, and A. Imamoglu, Signatures of a dissipative phase transition in photon correlation measurements, Nat. Phys. {\bf 14}, 365 (2018).

\bibitem{Rota-18} R. Rota, F. Minganti, A. Biella, and C. Ciuti, Dynamical properties of dissipative XYZ Heisenberg lattices, New J. Phys. {\bf 20}, 045003 (2018).

\bibitem{Ferreira-19} J. S. Ferreira and P. Ribeiro, Lipkin-Meshkov-Glick model with Markovian dissipation: A description of a collective spin on a metallic surface, Phys. Rev. B {\bf 100}, 184422 (2019).

\bibitem{Tomadin-11} A. Tomadin, S. Diehl, and P. Zoller, Nonequilibrium phase diagram of a driven and dissipative many-body system, Phys. Rev. A {\bf 83}, 013611 (2011).

\bibitem{Lee-11} T. E. Lee, H. H\"affner, and M. C. Cross, Antiferromagnetic phase transition in a nonequilibrium lattice of Rydberg atoms, Phys. Rev. A {\bf 84}, 031402(R) (2011).

\bibitem{Torre-13} E. G. D. Torre, S. Diehl, M. D. Lukin, S. Sachdev, and P. Strack, Keldysh approach for nonequilibrium phase transitions in quantum optics: Beyond the Dicke model in optical cavities, Phys. Rev. A {\bf 87}, 023831 (2013).

\bibitem{Lee-13} T. E. Lee, S. Gopalakrishnan, and M. D. Lukin, Unconventional Magnetism via Optical Pumping of Interacting Spin Systems, Phys. Rev. Lett. {\bf 110}, 257204 (2013).

\bibitem{Ludwig-13} M. Ludwig and F. Marquardt, Quantum Many-Body Dynamics in Optomechanical Arrays, Phys. Rev. Lett. {\bf 111}, 073603 (2013).

\bibitem{Carr-13} C. Carr, R. Ritter, C. G. Wade, C. S. Adams, and K. J. Weatherill, Nonequilibrium Phase Transition in a Dilute Rydberg Ensemble, Phys. Rev. Lett. {\bf 111}, 113901 (2013).

\bibitem{Sieberer-13} L. M. Sieberer, S. D. Huber, E. Altman, and S. Diehl, Dynamical Critical Phenomena in Driven-Dissipative Systems, Phys. Rev. Lett. {\bf 110}, 195301 (2013).

\bibitem{Sieberer-14} L. M. Sieberer, S. D. Huber, E. Altman, and S. Diehl, Nonequilibrium functional renormalization for driven-dissipative Bose-Einstein condensation, Phys. Rev. B {\bf 89}, 134310 (2014).

\bibitem{Marcuzzi-14} M. Marcuzzi, E. Levi, S. Diehl, J. P. Garrahan, and I. Lesanovsky, Universal Nonequilibrium Properties of Dissipative Rydberg Gases, Phys. Rev. Lett. {\bf 113}, 210401 (2014).

\bibitem{Weimer-15} H. Weimer, Variational Principle for Steady States of Dissipative Quantum Many-Body Systems, Phys. Rev. Lett. {\bf 114}, 040402 (2015).

\bibitem{Maghrebi-16} M. F. Maghrebi and A. V. Gorshkov, Nonequilibrium many-body steady states via Keldysh formalism, Phys. Rev. B {\bf 93}, 014307 (2016).

\bibitem{Sieberer-16} L. M. Sieberer, M. Buchhold, and S. Diehl, Keldysh field theory for driven open quantum systems, Rep. Prog. Phys. {\bf 79}, 096001 (2016).

\bibitem{Biondi-17} M. Biondi, G. Blatter, H. E. T\"ureci, and S. Schmidt, Nonequilibrium gas-liquid transition in the driven-dissipative photonic lattice, Phys. Rev. A {\bf 96}, 043809 (2017).

\bibitem{Domokos-17} J. M. Fink, A. Dombi, A. Vukics, A. Wallraff, and P. Domokos, Observation of the Photon-Blockade Breakdown Phase Transition, Phys. Rev. X {\bf 7}, 011012 (2017).

\bibitem{Young-20} J. T. Young, A. V. Gorshkov, M. Foss-Feig, and M. F. Maghrebi, Nonequilibrium Fixed Points of Coupled Ising Models, Phys. Rev. X {\bf 10}, 011039 (2020).

\bibitem{Mori-20}  T. Mori and T. Shirai, Resolving a Discrepancy between Liouvillian Gap and Relaxation Time in Boundary-Dissipated Quantum Many-Body Systems, Phys. Rev. Lett. {\bf 125}, 230604 (2020).

\bibitem{Haga-21} T. Haga, M. Nakagawa, R. Hamazaki, and M. Ueda, Liouvillian Skin Effect: Slowing Down of Relaxation Processes without Gap Closing, Phys. Rev. Lett. {\bf 127}, 070402 (2021).

\bibitem{Bensa-21} J. Bensa and M. {\v{Z}}nidari{\v{c}}, Fastest local entanglement scrambler, multistage thermalization, and a non-Hermitian phantom, Phys. Rev. X {\bf 11}, 031019 (2021).

\bibitem{Mori-23}  T. Mori and T. Shirai, Symmetrized Liouvillian Gap in Markovian Open Quantum Systems, Phys. Rev. Lett. {\bf 130}, 230404 (2023).

\bibitem{Diehl-08} S. Diehl, A. Micheli, A. Kantian, B. Kraus, H. P. B\"uchler, and P. Zoller, Quantum states and phases in driven open quantum systems with cold atoms, Nat. Phys. {\bf 4}, 878 (2008).

\bibitem{Kraus-08} B. Kraus, H. P. B\"uchler, S. Diehl, A. Kantian, A. Micheli, and P. Zoller, Preparation of entangled states by quantum Markov processes, Phys. Rev. A {\bf 78}, 042307 (2008).

\bibitem{Bonnes-14} L. Bonnes, D. Charrier, and A. M. L\"auchli, Dynamical and steady-state properties of a Bose-Hubbard chain with bond dissipation: A study based on matrix product operators, Phys. Rev. A {\bf 90}, 033612 (2014).

\bibitem{Griessner-06} A. Griessner, A. J. Daley, S. R. Clark, D. Jaksch, and P. Zoller, Dark-State Cooling of Atoms by Superfluid Immersion, Phys. Rev. Lett. {\bf 97}, 220403 (2006).

\bibitem{Pichler-13} H. Pichler, J. Schachenmayer, A. J. Daley, and P. Zoller, Heating dynamics of bosonic atoms in a noisy optical lattice, Phys. Rev. A {\bf 87}, 033606 (2013).

\bibitem{Stannigel-14} K. Stannigel, P. Hauke, D. Marcos, M. Hafezi, S. Diehl, M. Dalmonte, and P. Zoller, Constrained Dynamics via the Zeno Effect in Quantum Simulation: Implementing Non-Abelian Lattice Gauge Theories with Cold Atoms, Phys. Rev. Lett. {\bf 112}, 120406 (2014).

\bibitem{Cai-13} Z. Cai and T. Barthel, Algebraic versus Exponential Decoherence in Dissipative Many-Particle Systems, Phys. Rev. Lett. {\bf 111}, 150403 (2013).

\bibitem{Deng-10} H. Deng, H. Haug, and Y. Yamamoto, Exciton-polariton Bose-Einstein condensation, Rev. Mod. Phys. {\bf 82}, 1489 (2010).

\bibitem{Carusotto-13} I. Carusotto and C. Ciuti, Quantum fluids of light, Rev. Mod. Phys. {\bf 85}, 299 (2013).

\bibitem{Byrnes-14} T. Byrnes, N. Y. Kim, and Y. Yamamoto, Exciton-polariton condensates, Nat. Phys. {\bf 10}, 803 (2014).

\bibitem{Buca-12} B. Bu{\v{c}}a and T. Prosen, A note on symmetry reductions of the Lindblad equation: transport in constrained open spin chains, New J. Phys. {\bf 14}, 073007 (2012).

\bibitem{Wouters-06} M. Wouters and I. Carusotto, Absence of long-range coherence in the parametric emission of photonic wires, Phys. Rev. B {\bf 74}, 245316 (2006).

\bibitem{Wouters-07} M. Wouters and I. Carusotto, Excitations in a Nonequilibrium Bose-Einstein Condensate of Exciton Polaritons, Phys. Rev. Lett. {\bf 99}, 140402 (2007).

\bibitem{Szymanska-06} M. H. Szyma\'nska, J. Keeling, and P. B. Littlewood, Nonequilibrium Quantum Condensation in an Incoherently Pumped Dissipative System, Phys. Rev. Lett. {\bf 96}, 230602 (2006).

\bibitem{Znidaric-15} M. {\v{Z}}nidari{\v{c}}, Relaxation times of dissipative many-body quantum systems, Phys. Rev. E {\bf 92}, 042143 (2015).

\bibitem{Haga-23} T. Haga, M. Nakagawa, R. Hamazaki, and M. Ueda, Quasiparticles of decoherence processes in open quantum many-body systems: Incoherentons, Phys. Rev. Research {\bf 5}, 043225 (2023).

\bibitem{Lieb-72} E. Lieb and D. Robinson, The finite group velocity of quantum spin systems, Commun. Math. Phys. {\bf 28}, 251 (1972).

\bibitem{Poulin-10} D. Poulin, Lieb-Robinson Bound and Locality for General Markovian Quantum Dynamics, Phys. Rev. Lett. {\bf 104}, 190401 (2010).

\bibitem{footnotes} Several exceptions exist regarding the relation between relaxation time and the Liouvillian gap. Notably, the Liouvillian skin effect can lead to divergent relaxation times without necessitating the closure of the gap \cite{Mori-20, Haga-21, Bensa-21, Mori-23}. However, it is generally anticipated that the Liouvillian skin effect does not manifest in translation-invariant systems.

\bibitem{Iemini-18} F. Iemini, A. Russomanno, J. Keeling, M. Schir{\`o}, M. Dalmonte, and R. Fazio, Boundary Time Crystals, Phys. Rev. Lett. {\bf 121}, 035301 (2018).

\bibitem{Lledo-19} C. Lled\'o, Th. K. Mavrogordatos, and M. H. Szyma\'nska, Driven Bose-Hubbard dimer under nonlocal dissipation: A bistable time crystal, Phys. Rev. B {\bf 100}, 054303 (2019).

\bibitem{Piccitto-21} G. Piccitto, M. Wauters, F. Nori, and N. Shammah, Symmetries and conserved quantities of boundary time crystals in generalized spin models, Phys. Rev. B {\bf 104}, 014307 (2021).

\bibitem{Prazeres-21} L. F. Prazeres, L. S. Souza, and F. Iemini, Boundary time crystals in collective $d$-level systems, Phys. Rev. B {\bf 103}, 184308 (2021).

\bibitem{Kongkhambut-22} P. Kongkhambut, J. Skulte, L. Mathey, J. G. Cosme, A. Hemmerich, H. Ke{\ss}ler, Observation of a continuous time crystal, Science {\bf 377}, 670–673 (2022).

\bibitem{Hajdusek-22} M. Hajdu{\v{s}}ek, P. Solanki, R. Fazio, and S. Vinjanampathy, Seeding Crystallization in Time, Phys. Rev. Lett. {\bf 128}, 080603 (2022).

\bibitem{Krishna-23} M. Krishna, P. Solanki, M. Hajdu{\v{s}}ek, and S. Vinjanampathy, Measurement-Induced Continuous Time Crystals, Phys. Rev. Lett. {\bf 130}, 150401 (2023).

\bibitem{Derrida-98} B. Derrida, An exactly soluble non-equilibrium system: The asymmetric simple exclusion process, Phys. Rep. {\bf 301}, 65 (1998).

\bibitem{Hinrichsen-00} H. Hinrichsen, Non-equilibrium critical phenomena and phase transitions into absorbing states, Adv. Phys. {\bf 49}, 815 (2000).

\bibitem{Marchetti-13} M. C. Marchetti, J. F. Joanny, S. Ramaswamy, T. B. Liverpool, J. Prost, M. Rao, and R. A. Simha, Hydrodynamics of soft active matter, Rev. Mod. Phys. {\bf 85}, 1143 (2013).

\bibitem{Kleinherbers-20} E. Kleinherbers, N. Szpak, J. K\"onig, and R. Sch\"utzhold, Relaxation dynamics in a Hubbard dimer coupled to fermionic baths: Phenomenological description and its microscopic foundation, Phys. Rev. B {\bf 101}, 125131 (2020).

\end{thebibliography}
\end{document}